\documentclass[nonacm,table]{acmart}

\AtBeginDocument{%
  \providecommand\BibTeX{{%
    \normalfont B\kern-0.5em{\scshape i\kern-0.25em b}\kern-0.8em\TeX}}}

\usepackage{longtable}

\usepackage{amsmath,amssymb,amsfonts}
\usepackage{algorithm}
\usepackage[noend]{algpseudocode}
\usepackage{textcomp}
\usepackage{xcolor}
\usepackage{soul}
\usepackage{booktabs}
\usepackage[utf8]{inputenc}
\usepackage[T1]{fontenc}
\usepackage{microtype}
\usepackage{rotating}
\usepackage{graphicx}
\usepackage{paralist}
\usepackage{tabularx}
\usepackage{multicol}
\usepackage{multirow}
\usepackage{pbox}
\usepackage{enumitem}	
\usepackage{colortbl}
\usepackage{pifont}
\usepackage{xspace}
\usepackage{url}
\usepackage{tikz}
\usepackage{tabularx}
\usepackage{fontawesome}
\usepackage{lscape}
\usepackage{listings}
\usepackage{color}
\usepackage{anyfontsize}
\usepackage{comment}
\usepackage{soul}
\usepackage{multibib}
\usepackage{lmodern}
\usepackage{array}
\include{macro}

\newcommand{\revised}[1]{\textcolor{black}{#1}}
\newcommand{\rev}[1]{\textcolor{black}{#1}}
\newcommand{\manually}{551\xspace}
\newcommand{\commitsmined}{213,102\xspace}
\newcommand{\refactorings}{287,813\xspace}
\newcommand{\systems}{150\xspace}
\newcommand{\taxonomycats}{67\xspace}

\newcommand{\reftool}{RMiner\xspace}

\begin{document}

\title[Why Developers Refactor Source Code: A Mining-based Study]{Why Developers Refactor Source Code:\\ A Mining-based Study}

\author[J. Pantiuchina]{Jevgenija Pantiuchina}
\email{jevgenija.pantiuchina@usi.ch}
\affiliation{%
  \institution{Universit\`a della Svizzera italiana}
  \city{Lugano, Switzerland}
}

\author[F. Zampetti]{Fiorella Zampetti}
\email{fiorellazampetti@gmail.com}
\affiliation{%
  \institution{University of Sannio}
  \city{Benevento, Italy}
}

\author[S. Scalabrino]{Simone Scalabrino}
\email{simone.scalabrino@unimol.it}
\affiliation{%
  \institution{University of Molise}
  \city{Pesche, Italy}
}

\author[V. Piantadosi]{Valentina Piantadosi}
\email{valentina.piantadosi@unimol.it}
\affiliation{%
  \institution{University of Molise}
  \city{Pesche, Italy}
}

\author[R. Oliveto]{Rocco Oliveto}
\email{rocco.oliveto@unimol.it}
\affiliation{%
  \institution{University of Molise}
  \city{Pesche, Italy}
}

\author[G. Bavota]{Gabriele Bavota}
\email{gabriele.bavota@usi.ch}
\affiliation{%
  \institution{Universit\`a della Svizzera italiana}
  \city{Lugano, Switzerland}
}

\author[M. Di Penta]{Massimiliano Di Penta}
\email{dipenta@unisannio.it}
\affiliation{%
  \institution{University of Sannio}
  \city{Benevento, Italy}
}

\begin{abstract}
Refactoring aims at improving code non-functional attributes without modifying its external behavior. Previous studies investigated the motivations behind refactoring by surveying developers.
\rev{With the aim of generalizing and complementing their findings,} we present a large-scale study quantitatively and qualitatively investigating \emph{why} developers perform refactoring in open source projects. First, we mine \refactorings refactoring operations performed in the history of \systems systems. Using this dataset, we investigate the interplay between refactoring operations and process (\eg previous changes/fixes) and product (\eg quality metrics) metrics. Then, we manually analyze \manually merged pull requests implementing refactoring operations, and classify the motivations behind the implemented refactorings (\eg removal of code duplication). Our results led to (i) quantitative evidence of the relationship existing between certain process/product metrics and refactoring operations; and (ii) a detailed taxonomy, \rev{generalizing and }complementing the ones existing in the literature, of motivations pushing developers to refactor source code.
\end{abstract}


\maketitle

% !TEX root = main.tex

%%%%%%%%%%%%%%%%%%%%%%
%%%%%%%%%%%%%%%%%%%%%%
\section{Introduction} \label{sec:intro}
%%%%%%%%%%%%%%%%%%%%%%
%%%%%%%%%%%%%%%%%%%%%%

Software refactoring has been widely studied in the research community, with most of the works falling into three main research threads: (i) approaches aimed at identifying refactoring opportunities \cite{Palomba:survey}; (ii) techniques to recommend refactoring solutions for a given design flaw \cite{Bavota:refSurvey}; and (iii) empirical studies looking at software refactoring from many different perspectives \cite{Wang:icsm2009,Emerson:tse2011,Silva:fse2016,Cedrim:fse2017,Vassallo:scp2019}. 
The knowledge of motivations pushing developers to perform refactoring \cite{Silva:fse2016} 
can help in building recommender systems able to propose suitable solutions for that. For this reason, understanding when and why developers perform refactoring has been the goal of many previous studies \cite{Wang:icsm2009,Kim:fse2012,Bavota:jss2015,Silva:fse2016,Vassallo:scp2019}. 

Some of these studies tried to answer this question by looking at specific factors that might correlate with refactoring operations, such as code quality proxies \revised{(\ie quantitative measures providing indications about the internal quality of code components, such as quality metrics or code smells)} \cite{Bavota:jss2015,Vassallo:scp2019}. While valuable, these studies provide limited insights into the reasons behind the performed refactorings, since their analysis is mostly quantitative and limited to a small number of factors. Other studies opted for a more qualitative approach by interviewing developers \cite{Wang:icsm2009,Kim:fse2012} to identify the major factors that motivate their refactorings. Although these studies have pioneered the investigation of the reasons pushing developers to refactor their code, as observed by Silva \etal \cite{Silva:fse2016}, the previously mentioned surveys are general purpose, meaning that they do not ask developers to \emph{justify} specific refactorings they performed, but rather study refactoring habits in general. To address this limitation, Silva \etal \cite{Silva:fse2016} interviewed developers who authored 222 refactoring-related commits to understand the reasons behind these specific operations. 

Stemming from the studies discussed above \rev{and to generalize their findings \cite{Wang:icsm2009,Kim:fse2012,Silva:fse2016}}, this paper describes large-scale mining study combining quantitative and qualitative analyses to investigate the motivations behind refactoring operations, by observing code and discussions rather than interviewing developers.
From a quantitative point of view, we mine the change history of \systems Java repositories hosted on GitHub to extract \refactorings refactoring operations of 25 different types performed by developers through the \reftool tool \cite{Tsantalis:icse2018}. Then, we analyze product- (\eg slopes indicating whether the quality of code components as assessed by quality metrics is decreasing over time) and process-related (\eg source code change- and fault-proneness) factors that contribute to trigger refactoring actions. As compared to previous work~\cite{Bavota:jss2015,Vassallo:scp2019}, we consider a more comprehensive set of factors and, more importantly, analyze them in a single model rather than in isolation, showing which ones are related to refactoring operations. From a qualitative point of view, we use the same set of systems to manually analyze a statistically significant sample of \manually pull requests (PRs) in which (i) developers discuss refactoring \textbf{and} (ii) \reftool identifies at least one refactoring operation. Through a manual analysis, we identify the rationale of the refactoring change, and whether it is the main intent of the change or, rather, they are triggered by the code review process of the PR. As main contribution of this analysis, we defined an extensive taxonomy of \taxonomycats motivations pushing developers to implement refactoring operations. Our qualitative analysis complements and \rev{generalizes} the findings in previous survey-based studies \cite{Wang:icsm2009,Kim:fse2012,Silva:fse2016} by investigating the same research question with a completely different experimental design. 

As compared to the most similar work (\ie Silva \etal \cite{Silva:fse2016}), the following notable differences can be highlighted for what concerns the study design and findings:

\begin{itemize}
\item \emph{Study Design: surveying developers vs analyzing their activities}. While Silva \etal contacted the developers authoring the refactorings asking their motivations for the implemented changes, we manually inspect pull requests implementing refactorings by analyzing their discussion and related commits in order to create our taxonomy of motivations. Investigating the same research question with two different experimental designs can lead to additional insights, and \rev{helps in generalizing previous findings}.

\item \emph{Study Design: complementing qualitative and quantitative analysis}. In our work, we analyze the motivations behind refactoring operations not only from a qualitative perspective (as done by Silva \etal \cite{Silva:fse2016}), but also by quantitatively studying the influence of product and process metrics on the triggering of refactoring operations. Also, to the best of our knowledge, our study is the first one analyzing these metrics in a single model rather than in isolation (as done in \cite{Bavota:jss2015}, for example).

\item \emph{Findings: complementing and \rev{generalizing} Silva \etal \cite{Silva:fse2016}}. As output of their study, Silva \etal defined a list of 44 motivations for 12 frequently applied refactoring operations. Our taxonomy, besides confirming 41 of their motivations, \rev{thus improving the generalizability of their findings}, includes 26 additional ones that are not covered in the previous study. %Thus, a different study design allowed us to discover additional refactoring motivations, complementary to the ones previously known in the literature.
\end{itemize}

Our quantitative analysis indicates that code readability and process-related factors correlate with the changes a commit containing refactoring operations has. As the main result of the qualitative analysis, we provide a comprehensive taxonomy of \taxonomycats categories of motivations leading developers to refactoring operations. We describe and exemplify each category, and discuss its implications in refactoring research and practice.
% !TEX root = main.tex

\newcommand{\rqone}{\revised{Which product and process-related factors relate with an increase of refactoring operation chances?}}
\newcommand{\rqtwo}{What are the reasons for performing a refactoring operation?}

%%%%%%%%%%%%%%%%%%%%%%
%%%%%%%%%%%%%%%%%%%%%%
\section{Study Design} \label{sec:design}
%%%%%%%%%%%%%%%%%%%%%%
%%%%%%%%%%%%%%%%%%%%%%

The \emph{goal} of this study is to quantitatively and qualitatively analyze the context in which refactoring operations occur in open source projects, with the aim of identifying the circumstances that may make a refactoring happen. The \emph{quality focus} relates not only to code quality, but also to the improvement of the software development process. 
The \emph{context} consists of \refactorings refactoring actions automatically identified in \systems open-source projects and, for the qualitative analysis, of \manually manually-analyzed PRs mentioning refactoring operations and linked onto refactoring-related commits.

We address the following two research questions (RQs):

\textbf{RQ$_1$:} \emph{\rqone} We are interested in studying if various source code features or process features correlate with the presence of refactoring operations in a commit.

\textbf{RQ$_2$:} \emph{\rqtwo} We investigate the rationale behind refactoring opportunities. We consider refactorings occurred in PRs, and perform a qualitative analysis of developers' discussions over the PR. Also, since previous work found that most refactoring operations occur with other changes \cite{Emerson:tse2011}, by analyzing PRs we give a closer look at this phenomenon, investigating if the refactoring was tangled with other changes, and looking at whether the refactoring was the primary purpose of the PR. \revised{We decided to answer this RQ by looking at PRs rather than at commits implementing refactorings since PRs offer a richer set of information to analyze to derive the rationale behind refactoring operations. Indeed, they often feature a discussion among developers that can help in better understanding what the goal of the implemented change was.}\smallskip

\revised{The formulated RQs investigate the same phenomenon (\ie what the motivations for refactoring operations are) from two different perspectives (quantitative---RQ$_1$ \emph{vs} qualitative---RQ$_2$). The catalog of motivations identified in the two RQs can complement and support each other.}

\begin{figure*}[!h]
\begin{center}
\includegraphics[width=\linewidth]{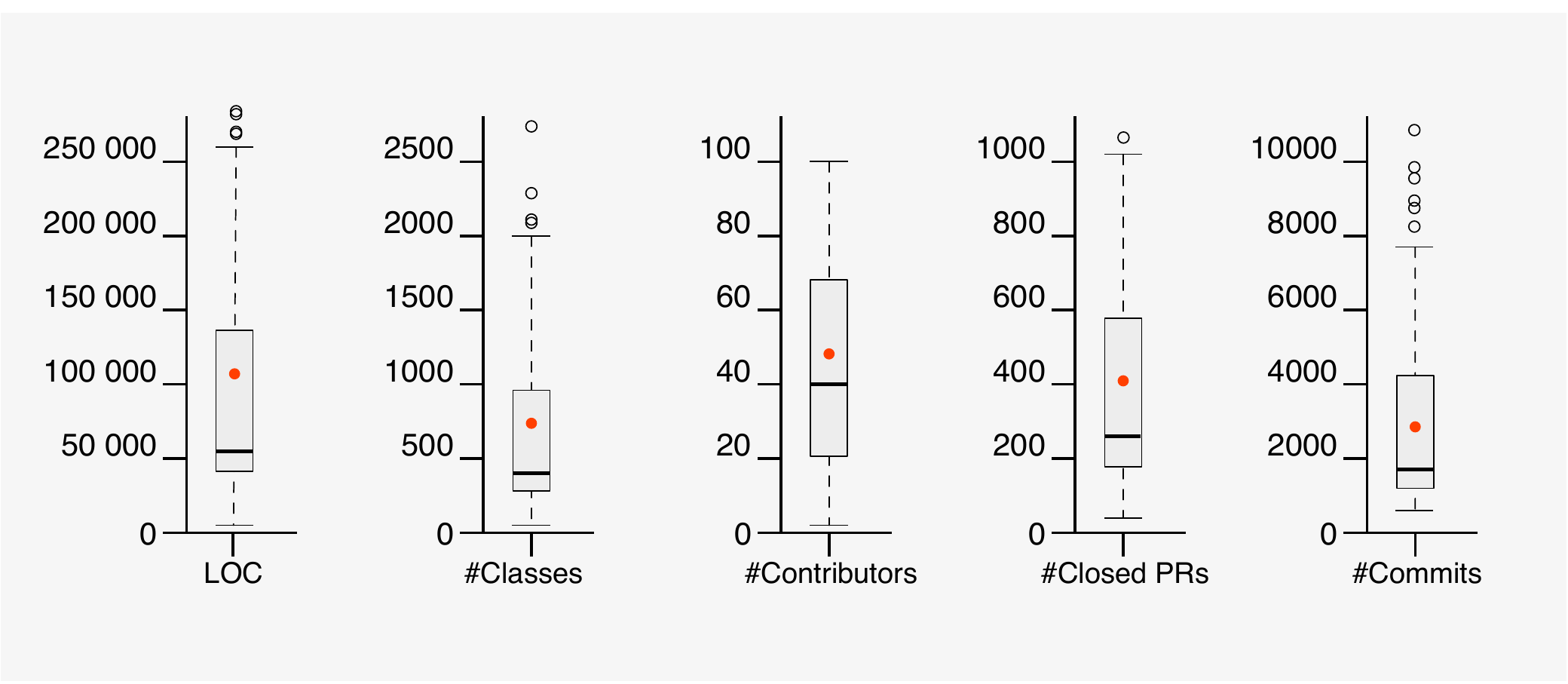}
\caption{\revised{Characteristics of the 150 projects used in our study}}
\label{fig:projects}
\end{center}
\end{figure*}

\subsection{Study Context}
We identified the projects to be studied among repositories hosted on GitHub. Since the infrastructure used in our study (\eg the refactoring detection tool) only supports Java, we focus on Java projects. \rev{Among all Java projects on GitHub, we aim at studying active projects having a non-trivial change history to study (needed to mine the PRs needed for our study) and not representing personal and/or toy projects (\eg a project created by a student during an assignment). To identify these projects we applied a number of selection criteria, only retaining projects having:
\begin{itemize}
\item \emph{At least 5 contributors and 1 fork}, to exclude personal/toy projects.
\item \emph{At least 500 commits and 100 PRs}, to exclude projects having a short change history and unlikely to provide useful PRs for our study.
\item \emph{Modified at least once in the period Jan-May 2019}, to exclude inactive projects at the time in which this study has been run.
\end{itemize}}

\revised{From the set of 303 remaining projects, we randomly selected 150 of them for our study (list available in \cite{replication}). The choice of selecting a subset of the 303 projects was dictated by the computationally expensive data extraction process adopted in our study. Indeed, as detailed in the following, besides detecting refactoring operations, we computed 42 product- and process-metrics (\eg code quality metrics, change-proneness of classes) for each of the 213,102 commits in the studied projects. This process took three months on a 56-core server.} \revised{\figref{fig:projects} reports boxplots depicting the distribution of Lines of Code (LOC), number of classes (\#Classes), number of contributors (\#Contributors), number of closed PRs (\#Closed PRs), and number of commits (\#Commits) for the analyzed 150 systems. The raw data from which this figure has been created is available in our replication package \cite{replication}.}

We used the \reftool tool~\cite{Tsantalis:icse2018} to detect the refactorings implemented by developers in the studied projects. We focus on commits performed in the master/default branch of each project. We have chosen \reftool due to its high reported precision (98\%) and recall (87\%)~\cite{Tsantalis:icse2018}. \reftool takes as input two consecutive commits and provides as output the set of detected refactorings (see \cite{Tsantalis:icse2018} for the supported refactorings).

\subsection{Quantitative Analysis (RQ$_1$)}
The occurrences of the detected refactorings constitute the \emph{dependent variable} for RQ$_1$. As \emph{independent variables}, we consider process-/product-related factors for each snapshot $s_i$ (commit) of the master branch. 

\subsubsection{Identification of product and process metrics}
The considered metrics are summarized in Tables \ref{tab:metrics}, \ref{tab:smells} and \ref{tab:process} and described in the following. The selection of these metrics (detailed in the following) is based on the will to include in our study:

\begin{enumerate}
\item \emph{Metrics capturing code quality from different perspectives (\tabref{tab:metrics}).} We included both structural and semantic (\ie textual) metrics that have been shown to capture orthogonal code quality aspects \cite{Marcus:icsm2005}. Also, we considered the recent readability metrics proposed in the literature \cite{Buse:tse2010,Scalabrino:ICPC16}, that have been shown to highly correlate with the developers' assessment of code readability.

\item \emph{Code smells and quality issues widely studied in the literature (\tabref{tab:smells}).} The presence of code smells has been correlated with higher change- and fault-proneness of code \cite{Palomba:emse2018} and, thus, they could also be responsible for the triggering of refactoring actions. Also, static analysis tools are more and more used in the context of continuous integration to perform basic code quality checks at commit time. Thus, we decided to include the warnings raised by one of these state-of-the-art tools, \ie PMD \cite{pmd}.

\item \emph{Process-related factors (\tabref{tab:process}).} These metrics are meant to provide a view on the development process, the developers involved in it, and historical information about the code components. We conjecture that these factors can play an important role in taking refactoring decisions, as also partially confirmed by previous work in the literature \cite{Vassallo:scp2019}.

\end{enumerate}

\revised{As detailed in \secref{sec:preprocess}, to avoid multicollinearity, we performed a variable selection.  Near each metric, we indicate the cluster it belongs to and whether (\checkmark) it was retained.} 

\textbf{Source Code Quality Metrics.} We consider, for each class $C$ changed in each snapshot $s_i$, its quality \emph{trend} as assessed by the 18 metrics in \tabref{tab:metrics}. These metrics capture different aspects of code quality, including size (\eg ELOC), coupling (\eg CBO), inheritance (\eg DIT), complexity (WMC), encapsulation (\eg NOPM), and readability (\eg StrRead). The first 13 metrics in \tabref{tab:metrics} (\ie until NOSI included) have been computed by using the CK tool~\cite{ck}. For the HsLCOM and C3, we used our implementation, \revised{while for the readability metrics we relied on the original implementations of the tools computing these metrics kindly made available by the original authors of the papers that introduced them~\cite{Buse:tse2010,Scalabrino:ICPC16}.}
 We start by measuring these 18 metrics on each class in each mined snapshot. Then, based on this information, we compute, for each snapshot, the slope of each metric over a window of N preceding commits (we set N=10 according to a previous work recommending just-in-time refactoring~\cite{Pantiuchina:icpc2018}). The slope of a line describes its steepness and in our case can highlight, for example, continuing degradation of some quality aspects (\eg a high positive slope for the WMC metrics indicates a steep increase in complexity for a class over time). Thus, using slopes we capture the improvement or degradation of quality factors, where the latter may trigger a refactoring. Clearly, slopes were considered unavailable for the first ten commits of a class.
 
 %$s_i$ the recent changes in code quality (\ie the ones resulting from commits preceding $s_i$) for a class $C$ as assessed by the 18 metrics. For each metric $M$ and class $C$ we compute its recent code quality trend, meaning the slope of the regression line fitting $M$'s values for $C$ in the ten commits preceding $s_i$. The slope of a line describes its steepness and in our case can highlight, for example, continuing degradation of some quality aspects (\eg a high positive slope for the WMC metrics indicates a steep increase in complexity for $C$ over time). Our goal is to find out whether developers tend to refactor more classes with recent worrying trends of some quality metrics. The ``ten'' threshold has been set accordingly to previous work in the literature~\cite{Pantiuchina:icpc2018}. Clearly, slopes were considered unavailable for the first ten commits of a class.

\newcommand{\SB}[1]{\textbf{#1} \checkmark}
\begin{table}
	\caption{Quality Metrics (Product-related factors). \revised{Near each factor, we indicate  whether (\checkmark) it was retained. The factors retained in the model are also highlighted in boldface.}}
	\label{tab:metrics}
	\footnotesize
		\rowcolors{2}{gray!15}{white}
		%\resizebox{\linewidth}{!}{
	\begin{tabular}{lp{100mm}l}
		\toprule
		\normalsize \bf Metric & \normalsize \bf Description \\
		\midrule
		  \SB{CBO}          &  Coupling Between Object classes: measures the dependencies a class has~\cite{Chidamber:tse1994}\\
		  \SB{WMC}          &  Weighted Methods per Class: sums the cyclomatic complexity of the methods in a class~\cite{Chidamber:tse1994}\\
		 \SB{RFC}     &  Response For a Class: the number of methods in a class plus the number of remote methods that are called recursively through the entire call tree~\cite{Chidamber:tse1994}\\
		  \SB{ELOC}         &  Effective Lines Of Code: the lines of code excluding blank lines and comments\\
		  \SB{NOM}          &  Number Of Methods in a class\\
		  \SB{NOPM}         &  Number Of Public Methods in a class\\
		 \SB{DIT}     &  Depth of Inheritance Tree: the length of the path from a class to its farthest ancestor~\cite{Chidamber:tse1994}\\
		%  \SB{LCOM}         &  Lack of Cohesion Of Methods: a class cohesion metric based on the sharing of local instance variables by the methods of the class~\cite{Chidamber:tse1994}\\
		\SB{NOC}     &  Number Of Children (direct subclasses) of a class\\
		\SB{NOF}     &  Number Of Fields declared in a class\\
		 \SB{NOSF}         &  Number Of Static Fields declared in a class\\
		 NOPF         &  Number Of Public Fields declared in a class\\
		\SB{NOSM}    &  Number Of Static Methods in a class\\
		\SB{NOSI}    &  Number Of Static Invocations of a class\\
		\SB{HsLCOM}  &  Henderson-Sellers revised Lack of Cohesion Of Methods (LCOM): a class cohesion metric based on the sharing of local instance variables by the methods of the class~\cite{Chidamber:tse1994}. HsLCOM dresses limitations of the original LCOM~\cite{HendersonSellers1996CouplingAC}\\
		\SB{C3}      &  Conceptual Cohesion of Classes: avg. textual similarity between all pairs of methods in a class \cite{Marcus:icsm2005}\\
		\SB{StrRead} &  Structural readability: uses structural aspects (\eg line length) to model code readability~\cite{Buse:tse2010} \\
		\SB{ComRead} &  Comprehensive readability model: combines structural, visual (\eg alignment) and textual features (\eg comments readability)~\cite{Scalabrino:ICPC16} \\
		\bottomrule	
	\end{tabular}
	%}
\end{table}

\textbf{Code Design Flaws and Quality Warnings.} We consider code design flaws related to the lack of adoption of good Object-Oriented coding practices (\ie Spaghetti Code, Excessive Coupling), to complex/large code components (\ie Blob Class, Complex Class) as well as other design flaws and warnings (\ie Excessive Imports, Too Many Methods) raised by a static analysis tool. We detect five types of code smells using an implementation of the DECOR smell detector based on the original rules defined by Moha \etal~\cite{Moha:tse2010}. The choice of using DECOR is driven by the fact that (i) it is a state-of-the-art smell detector having high accuracy in detecting smells~\cite{Moha:tse2010}; and (ii) it applies simple detection rules that allow it to be very efficient. The latter was a strict requirement for our analysis since we detected smells in all classes and for all studied systems' snapshots. In addition, we also consider 13 flaws from a widely-used static analysis tool that does not require code compilation, \ie PMD \cite{pmd}. The set of detected design flaws and code quality warnings is described in \tabref{tab:smells}.

\begin{table}
	\caption{Code Design Flaws (Product-related factors). \revised{Near each factor, we indicate  whether (\checkmark) it was retained. The factors retained in the model are also highlighted in boldface.}}
	\label{tab:smells}
	\footnotesize
	\centering
	\rowcolors{2}{gray!15}{white}
	%\resizebox{0.98\linewidth}{!}{
	\begin{tabular}{m{45mm}p{70mm}}
        \toprule
        \normalsize \bf Design Flaw & \normalsize \bf Description \\
        \midrule
        \multicolumn{2}{l}{\small \textbf{DECOR Code Smells}} \\
        Blob Class               & A large class that monopolizes most of the application logic~\cite{Brown98-AntiPatterns}\\
        \SB{Complex Class}                 & A class characterized by a high cyclomatic complexity~\cite{Brown98-AntiPatterns} \\
        \SB{Spaghetti Code}           & A class declaring long methods without parameters~\cite{Brown98-AntiPatterns}\\
        \SB{CDSBP}                    &  Class Data Should be Private: violation of information hiding principle~\cite{Fowler:1999}\\
        \SB{Functional Decomposition} & Scarcely used object-oriented principles, such as inheritance and polymorphism; few methods and many private fields~\cite{Brown98-AntiPatterns}\\
        \midrule
        
        \multicolumn{2}{l}{\small \textbf{PMD code quality warnings}} \\
        \SB{Excessive Coupling}                & A highly coupled class hindering reuse and maintainability~\cite{Fowler:1999}\\
        \SB{Too Many Nested If \mbox{Statements}}     & Makes the code harder to understand and increase error-proneness\\
         \SB{Excessive Imports}                      & It might indicate too high coupling\\ 
         \textbf{Too High NPath} \mbox{Complexity}         & NPath is the number of acyclic execution paths throughout a method\\ 
        \SB{Excessive Method Length}                & It might indicate too many functionalities in a single method\\ 
         Excessive Class Length                 & It might indicate too many responsibilities implemented in a class\\ 
         \SB{Too Many Fields}                        & It can make the code hard to understand\\
         \SB{Too Many Methods}                       & It might indicate too many responsibilities in a class\\
         Cyclomatic Complexity                  & An excessive degree of decisional logic in a class\\
         \textbf{Excessive Parameter} \mbox{\SB{List}} & It might indicate the need for a new object to wrap them\\ 
         \SB{NCSS Type Count}                        & Similar to excessive class length, but it only considers actual statements\\ 
         \SB{NCSS Method Count}                      & Similar to excessive method length, but it only considers actual statements \\
         \SB{NCSS Constructor Count}                 & Equivalent of NCSS Method Count  for constructors\\
		\bottomrule
	\end{tabular}
	%}
\end{table}

\textbf{Process-related factors.} Besides the product-related factors previously described, we also study how process-related factors correlate with refactoring. In this case, we extract for each analyzed snapshot the factors summarized in \tabref{tab:process}.  

Given a snapshot $s_i$, we compute its distance (in commits) from the previous and next release
  (first two rows in \tabref{tab:process}). This to verify the conjecture of Vassallo \etal~\cite{Vassallo:scp2019} that refactoring does not occur immediately before/after a release.
This information was retrieved using the GitHub API, through which it is possible to access all the tags related to a project. Then, we manually looked at the tags assigned to each project to isolate the ones referring to a new release.

We also consider the change- and fault-proneness of classes. The change-proneness is computed as the ratio between the total number of lines changed in the class $C$ from the date of its addition to the project and the total number of commits in which $C$ was changed, until each snapshot $s_i$. 

The fault-proneness for $C$ is computed as the number of bug-fixing commits it has been subject to in the past (\ie before $s_i$). For each project, we firstly identified all bug-fixing commits by matching patterns~\cite{DBLP:conf/icsm/FischerPG03}:  ``fix'' or ``solve'' or ``close'' \textbf{and} ``bug'' or ``defect'' or ``crash'' or ``fail'' or ``error''.  Then, for a given class $C$ and for each snapshot $s_i$, we compute the number of bug-fixing commits preceding $s_i$ and impacting $C$. Section \ref{sec:threats} discusses the extent to which this simple heuristic for identifying bug fixes leads towards imprecisions.

Finally, we consider two metrics capturing the experience of the developers who worked on the system's classes. The first metric, named \emph{Developer Overall Experience}, assesses the experience of each developer as the number of commits she performed in the past. For each snapshot $s_i$ and for each of its classes $C$, we extract the list of developers who modified $C$ in the past (\ie before $s_i$). For each commit $c_j$ (with $j<i$) in which $C$ has been modified, we compute the experience of the developer authoring $c_j$ (\ie the number of commits she performed \emph{before} $c_j$). This gives us a distribution of developers' experiences, for which we compute the minimum. Indeed, the minimum represents the lowest experience of a developer who worked on $C$, and we assume it might be correlated with future refactoring actions taken on $C$.

The \emph{Developer Class Experience} computes a class-related experience: for each snapshot $s_{i}$ and for each of its classes $C$, this form of experience is computed for a given developer as the number of commits impacting $C$ she performed in the past. Thus, it is a more specific version of the overall experience. We compute this metric for each s$_i$ and $C$ under study in the same way explained for the overall experience.

\begin{table}
	\caption{Process-related factors. \revised{Near each factor, we indicate  whether (\checkmark) it was retained. The factors retained in the model are also highlighted in boldface.}}
	\label{tab:process}
	\footnotesize
	\centering
	\rowcolors{2}{gray!15}{white}
	%\resizebox{0.96\linewidth}{!}{
	\begin{tabular}{p{45mm}p{70mm}}
		\toprule
		\normalsize \bf Metric & \normalsize  \bf Description \\
		\midrule
		%\multirow{2}{1.5cm}{Closeness to a previous bug fix release} & TODO \newline TODO\\
		\SB{Closeness to a previous release}  & The number of commits until the previous minor/major release\\
		\SB{Closeness to a next release}      & The number of commits until the next minor/major release\\
		\SB{Fault-Proneness}                  & Number of bugs fixed in the project history on a given class\\
		\SB{Change-Proneness}                 & The average number of lines impacted in commits related to a class \\
		\SB{Developer Overall Experience}     & The number of past commits a developer performed \\
		\SB{Developer Class Experience}       & The number of past commits on a class performed by a developer\\
		\bottomrule
	\end{tabular}
	%}
\end{table}

\subsubsection{Metrics Aggregation and Preprocessing}
\label{sec:preprocess}
Since in RQ$_1$ we are interested to build an \textit{explanatory model} explaining which factors correlate with the presence of refactoring actions in a snapshot, we had to aggregate metrics for all classes involved in each snapshot. For the product metrics, we compute the maximum slope among all classes involved in the snapshot, except for the conceptual cohesion (\textit{C3}) and readability (\textit{StrRead} and \textit{ComRead}) metrics, which go in the opposite directions than other metrics (higher values are better). In such cases, we consider the minimum. In both cases, the rationale is to identify the ``worst case" in a snapshot, which could ideally trigger a refactoring.
As for the DECOR smells, we count the number of classes exhibiting a smell in each snapshot, while for PMD we sum the number of warnings of each type among changed classes.
Similarly to what done for product metrics, for process metrics, we compute the maximum (\eg maximum number of bugs), except for the experience-related metrics, where we consider the minimum, again to consider the worst-case scenario. Finally, release-related metrics do not need to be aggregated, since they are already at commit granularity.

\revised{After that, to avoid multi-collinearity, we use the {\em R redun} function of the {\em Hmisc} package~\cite{hmiscr} for removing redundant variables. The {\em redun} function stepwise removes variable, starting from the most predicable one, until no variable can be predicted with an adjusted $R^2$ greater than a given threshold (0.8 in our study).}
%produce a hierarchical clustering of features based on their correlation, in turn, computed with a specified correlation measure (we use the Spearman's $\rho$ rank correlation). Then, we identify clusters by cutting the tree at a given level of $\rho^2$ that we set at $\rho^2=0.25$, which corresponds to a medium correlation (\ie $\rho=0.5$)~\cite{Cohen-1988}. 
Once again, we use the whole dataset to perform correlation analysis, because we intend to build an explanatory model and not a predictive model.

Since the value of our independent variables can depend on projects' characteristics, and to properly interpret the importance of each variable in the model, we normalize variable values, within each project, in the interval [0, 1]. \revised{This is done by subtracting the minimum and dividing by the difference between the maximum and minimum.}
Finally, to build a model easy to be interpreted, we invert (\ie compute 1-x) the values of variables going towards a different direction than the others (\ie those for which the higher the better).

\subsubsection{Mixed-model Building}
Once variables have been preprocessed, we address \textbf{RQ$_1$} by building mixed-effect generalized linear models. The model, built using the \emph{glmer} function of the \emph{lme4} \cite{lme4} \tool{R} package, is a logistic regression mixed-effect model where: (i) the dependent variable is a dichotomous variable indicating whether at least a refactoring was performed in a given commit; (ii) the independent variables (fixed effects) are all the aforementioned ones, after having pruned out those highly correlating with others; (iii) the random effect is the project in which the change occurred. The latter aims at controlling within-project effects, \eg a project following a specific development process had better code quality assurance policies than others. \revised{To simplify, our model reports whether the status of the system (as assessed by the used independent variables) in the snapshot $S_{i-1}$ triggered a refactoring in the subsequent commit $C_i$.}

To answer {\bf RQ$_1$}, we report the details of the model, among others the coefficient of each factor in the model, and the \emph{p}-value indicating whether the factor is statistically significant or not (for a significance level of 95\%). We also report the odds ratio (OR) which, for a logistic regression model, is given by $e^{c_i}$ where $c_i$ is the coefficient of the $i$-th factor. An OR $>1$ indicates that a unity increase of a variable increases of OR times the chances of a refactoring to occur.

\subsection{Qualitative Analysis of Refactoring Discussions in Pull Requests (RQ$_2$)}
For the qualitative analysis, we identified PRs likely discussing refactorings using two criteria to be satisfied: (i) whether a commit is a part of PR or made during its review contains a refactoring identified by \reftool, and (ii) whether the PR title or comments contain refactoring-related keywords. We used a list of refactoring keywords defined in a previous work \cite{Canfora:2014} \revised{(available in our replication package \cite{replication})} and augmented it with all names of refactorings identified by \reftool \cite{Tsantalis:icse2018}. Note that, while this selection process can generate false positives (\ie PRs unrelated to refactoring operations), these will be discarded during the manual analysis and, thus, do not represent a source of noise for our study.

Once the candidate set of 2,400 PRs has been identified, we created a randomly-stratified sample of \manually PRs. The strata here were represented by the projects, \ie PRs were sampled across projects  proportionally based on the number of candidate PRs found in the previous step. The total number of PRs sampled allows us to ensure a significance interval (margin of error) of $\pm5\%$ with a confidence level of 99\%, \revised{and feature a total of 8,108 refactoring operations identified by \reftool}. 
This estimation has been performed using a sample size ($SS$) calculation formula for an unknown population \cite{Rosner2011}:
\[
SS=p \cdot \left(1-p\right)\frac{Z_\alpha^2}{E^2}
\]

\noindent and $SS_{adj}$ for a known population $pop$:
\[
SS_{adj}=\frac{SS}{1+\frac{SS-1}{pop}}
\]

\noindent where $p$ is the estimated probability of the observation event to occur (we assume it being 0.5 if we don't know it a priori), $Z_\alpha$ is the value of the $Z$ distribution for a given confidence level, and $E$ is the estimated margin of error (5\%).

We then uploaded the sample of PRs on a tagging webapp we used to perform a manual coding of PRs. The webapp presented to the annotator the following information: (i) the PR title and hyperlink to the discussion; (ii) the refactoring-related keyword(s) matched in the PR text; and (iii) the list of refactorings detected by \reftool in commits linked to the PR, as well as the links to the GitHub diff pages of the commits themselves.

Through the coding app, each annotator could add one or more tagging items, containing the following information: (i) the type of refactoring action performed and discussed in the PR, or whether the change discussed was related to a combination of refactorings; (ii) whether the refactoring was the original intent of the PR, whether it happened as a consequence of the PR discussion, or whether it happened accidentally because of another change; (iii) whether the refactoring was tangled with other changes, or if it was the only purpose of the PR; (iv) finally, a tag indicating the motivation behind the refactoring, as it could be inferred from \revised{the inspection of the PR title/description, from its discussion, and from the commits related to it, looking at commit messages and, when needed, code diff}. Note that each annotator could add more than one motivation for each PR (\eg one for each refactoring operation, or even more than one for the same refactoring). To assign the tag describing the motivation, the annotator could choose an available tag in a drop-down menu (from those previously created by other annotators or by herself), or add a new one if no tag was fitting the specific case. If an annotator realized that the PR discussion was not related to refactoring, the PR was tagged as ``false positive''.

\revised{Six of the seven authors took part in the annotation process. The webapp we developed took care of automatically assigning each PRs to at least two of the involved annotators. We collected a total of 1,223 tags each one reporting a motivation for a refactoring (or combination of refactorings) performed in a PR. After each PR was tagged by two annotators, three of the authors jointly worked on the available tags to perform a card sorting activity \cite{cardSorting} aimed at merging duplicates (\ie similar tags having the same meaning), and start grouping tags into categories. Then, they created a first taxonomy describing the different purposes of refactorings by only using the 699 tags for which there was no conflict (\ie the same tag was used by the two annotators for motivating the refactoring observed in a PR). After a first draft of the taxonomy was produced, two different authors refined it, by renaming some categories and moving sub-categories through the taxonomy. Once the final taxonomy was produced, three authors jointly discussed the conflicting cases in the categorization (524 out of 1,223 tags)  and assigned them to suitable taxonomy categories, creating new ones when needed, and ensuring a consistency of category naming.}

To address \textbf{RQ$_2$}, we report and discuss the taxonomy of refactoring motivations inferred as previously explained. In particular, we discuss the various categories, highlighting the percentages of PRs belonging to the category, reporting some examples, and highlighting the implications resulting from our empirical findings.

%On top of that, we show for different types of refactoring operations detected in the PRs, the percentage of cases for which the refactoring action was the only purpose of the PR, as well as whether the refactoring was the original intent of the PR or, rather, it was triggered after the PR discussion.

% !TEX root = main.tex

%%%%%%%%%%%%%%%%%%%%%%
%%%%%%%%%%%%%%%%%%%%%%
\section{Results} \label{sec:results}
%%%%%%%%%%%%%%%%%%%%%%
%%%%%%%%%%%%%%%%%%%%%%

In the following we report and discuss the results addressing our RQs (\secref{sec:design}).

\subsection{\rqone}

Over the \commitsmined snapshots analyzed, \reftool identified a total of \refactorings refactoring operations. More in details, our dataset contains 35,560 commits ($\simeq 17\%$) with at least one refactoring operation. If we exclude renaming operations (\textit{Rename Method} and \textit{Rename Class}), \reftool found a total of 209,385 refactorings in 28,716 different snapshots (14\%).

\begin{table}[t]
	\caption{Generalized mixed effect logistic regression model: diagnostics, residuals, and random effect}
	\label{tab:rq1-diag}
	\centering
	%\resizebox{\linewidth}{!}{
	\begin{tabular}{llrrrrr}
		\toprule
		\multicolumn{7}{c}{\bf Diagnostics}\\
		\midrule
		&&     \textbf{AIC} &     \textbf{BIC} &  \textbf{logLik }& \textbf{deviance} &\textbf{df resid.}\\ 
		&&  21,071.5 & 21,362.4 & -10,499.7  & 20,999.5  &  23,860 \\ 
		\midrule
		\multicolumn{7}{l}{\bf Scaled residuals}\\ 
		\midrule
		&&\textbf{Min }  &    \textbf{1Q } &  \textbf{Median }   &   \textbf{3Q }  &   \textbf{Max}\\ 
		&&-2.0565  & -0.5256  & -0.3867  & -0.1094  & 11.3848 \\
		\midrule
		\multicolumn{7}{c}{\bf Random effects}\\
		\midrule
		\multicolumn{2}{l}{\bf Groups      Name}        & \textbf{Variance} &\textbf{Std.Dev.}\\
		\multicolumn{2}{l}{ProjectName (Intercept)} & 1.549  &  1.245   \\	
		\bottomrule
	\end{tabular}
\end{table}

\begin{table}[t]
	\caption{Generalized mixed effect logistic regression model: effect of considered factors}
	\label{tab:rq1}
	\centering
	%\resizebox{\linewidth}{!}{
	\begin{tabular}{llrrrrr}
		\toprule
		% \midrule
		% 	\multicolumn{6}{l}{\textbf{Quality Metrics}}\\		
% 		\midrule
		& \textbf{Metric}      & \textbf{OR}    & \textbf{Estimate} & \textbf{Std. error} & \textbf{\emph{z}-value} & \textbf{\emph{p}-value} \\ 
		\midrule
		& (Intercept) & 0.00 & -7.52 & 0.52 & -14.50 & \textbf{$<$0.01}  \\ 
		% \midrule
		\midrule
		\multirow{16}{*}{\rotatebox[origin=c]{90}{\textbf{Quality}}}				
		& \textbf{LackStructRead} & \textbf{3.14} & 1.14 & 0.38 & 3.05 & \textbf{$<$0.01}\\ 
		& \textbf{LackComRead }& \textbf{2.68} & 0.99 & 0.41 & 2.42 & \textbf{0.02} \\ 
		& \textbf{LackC3} & \textbf{1.87} & 0.63 & 0.30 & 2.06 & \textbf{0.04} \\ 
		& CBO & 1.02 & 0.02 & 0.03 & 0.66 & 0.51 \\ 
		& WMC & 0.98 & -0.02 & 0.01 & -1.57 & 0.12 \\ 
		& DIT & 1.17 & 0.16 & 0.12 & 1.39 & 0.17 \\ 
		& NOC & 0.95 & -0.06 & 0.28 & -0.20 & 0.84 \\ 
		& RFC & 0.98 & -0.02 & 0.02 & -1.05 & 0.29 \\ 
		& \textbf{NOM} & \textbf{0.88} & -0.12 & 0.05 & -2.54 & \textbf{0.01} \\ 
		& \textbf{NOPM} & \textbf{1.22} & 0.20 & 0.06 & 3.19 & \textbf{$<$0.01} \\ 
		& NOSM & 0.91 & -0.09 & 0.12 & -0.76 & 0.45 \\ 
		& NOF & 1.12 & 0.11 & 0.08 & 1.34 & 0.18 \\ 
		& NOSF & 0.89 & -0.11 & 0.13 & -0.88 & 0.38 \\ 
		& NOSI & 1.01 & 0.01 & 0.07 & 0.12 & 0.90 \\ 
		& LOC & 1.00 & 0.00 & 0.00 & 0.91 & 0.36 \\ 
		& \textbf{HsLCOM} & \textbf{1.94} & 0.66 & 0.29 & 2.28 & \textbf{0.02} \\ 	
		%   \midrule
		%   \multicolumn{6}{l}{\textbf{Code Design Flaws Metrics}}\\
		\midrule
		\multirow{12}{*}{\rotatebox[origin=c]{90}{\textbf{Code Design Flaws}}}	
		& IsGodDecor & 1.12 & 0.12 & 0.14 & 0.81 & 0.42 \\ 
		& IsCDSBPDecor & 0.90 & -0.11 & 0.15 & -0.71 & 0.48 \\ 
		& IsComplexDecor & 1.03 & 0.03 & 0.14 & 0.23 & 0.82 \\ 
		& IsFuncDecDecor & 0.76 & -0.28 & 0.25 & -1.11 & 0.27 \\ 
		& IsSpaghCodeDecor & 1.10 & 0.09 & 0.13 & 0.69 & 0.49 \\ 
		& AvoidDeeplyNestedIfStmts & 0.92 & -0.09 & 0.20 & -0.43 & 0.67 \\ 
		& CouplingBtwObjects & 0.73 & -0.31 & 0.22 & -1.42 & 0.16 \\ 
		& ExcessiveImport & 1.04 & 0.04 & 0.20 & 0.18 & 0.86 \\ 
		& ExcessiveMethodLength & 0.96 & -0.04 & 0.23 & -0.18 & 0.85 \\ 
		& ExcessiveParameterList & 1.28 & 0.25 & 0.20 & 1.22 & 0.22 \\ 
		& TooManyFields & 1.08 & 0.08 & 0.20 & 0.40 & 0.69 \\ 
		& TooManyMethods & 0.66 & -0.42 & 0.29 & -1.42 & 0.16 \\ 	
		%     \midrule
		%     \multicolumn{6}{l}{\textbf{Process-related Metrics}}\\
		\midrule
		\multirow{6}{*}{\rotatebox[origin=c]{90}{\textbf{Process}}}
		& \textbf{LackGeneralExp} & \textbf{1.26} & 0.23 & 0.10 & 2.30 & \textbf{0.02} \\ 
		& \textbf{LackFileExp} & \textbf{8.93} & 2.19 & 0.13 & 16.45 & \textbf{$<$0.01}  \\ 
		& \textbf{FilesRelatedToIssueFix} & \textbf{2.09} & 0.74 & 0.07 & 10.12 & \textbf{$<$0.01}  \\ 
		& \textbf{AvgLinesImpactedInCommit} & \textbf{1.93} & 0.66 & 0.17 & 3.96 & \textbf{$<$0.01}  \\ 
		& DistancePreviousRelease & 1.13 & 0.12 & 0.08 & 1.59 & 0.11 \\ 
		& \textbf{DistanceNextRelease} & \textbf{1.43} & 0.36 & 0.09 & 4.12 & \textbf{$<$0.01}  \\ 
		%     \midrule
		%     \multicolumn{7}{c}{\textsc{Random Effects}}\\ 
		% 	\midrule
		% 	
		% 	& \multicolumn{2}{l}{Name} & \multicolumn{2}{l}{\textbf{}} & \multicolumn{2}{l}{Std. Deviation}\\ 
		% 	& \multicolumn{2}{l}{\textbf{Project}} & \multicolumn{2}{l}{2.108} & \multicolumn{2}{l}{1.452}\\
		\bottomrule
	\end{tabular}
	%}
\end{table}

\revised{\tabref{tab:rq1-diag} and \tabref{tab:rq1} reports the results of the logistic regression mixed-effect model. More specifically, \tabref{tab:rq1-diag} reports the model diagnostics (Akaike Information Criterion --- AIC \cite{aic}, Bayesian Information Criterion --- BIC, log likelihood, deviance, and degree of freedom residuals), the scaled residuals, and the random effect (project estimate).
Concerning the model fitting (\tabref{tab:rq1-diag}), we tried different models, namely logistic (\ie the one reported, AIC=21,071), linear (AIC=22,565), and Poisson (AIC=22,560). Also, although the analysis performed using the \emph{redun} function already used a goodness-of-fit to iteratively remove variables, we experimented logistic models using structural metrics only (AIC=89,751), conceptual metrics only (AIC=89,475), code design flaws only (AIC=179,528), and process metrics only (AIC=23,583). Ultimately, the comprehensive logistic regression model we report is the one with the smallest AIC among those considered.}

\revised{\tabref{tab:rq1} reports the OR, estimate, standard error, \emph{z}-value and \emph{p}-value for the various factors we considered. We report in bold face the coefficient for which there is a  statistically significant correlation. Metrics that have been inverted (\eg C3) are named with the prefix ``Lack''.}

 %The estimated coefficient for the only random effect we considered (\ie project) is 2.11 ($\mathit{SD} = 1.45$).

\revised{Looking at code quality metrics, we found that the lack of structural readability plays a significant role: } lack of structural readability~\cite{Buse:tse2010} increases the odds of refactoring (OR=$3.14$). \revised{At the same time, \textit{ComRead} readability metric and \textit{LackC3} show a marginal significance \textit{(p-value} = 0.02 and \textit{p-value} = 0.04), respectively. In particular, looking at} the \textit{ComRead} readability metric combining structural and textual features~\cite{Scalabrino:ICPC16} the OR is $2.68$, while classes showing a decrease in their conceptual cohesion (\textit{LackC3}) have 1.87 times higher odds of being refactored. 

\revised{Among the structural metrics, we found that \textit{NOM}, \textit{NOPM} and \textit{HsLCOM} have a statistically significant effect (marginally significant for \textit{HsLCOM}), although the OR for \textit{NOM} and \textit{NOPM} is close to one. Instead, the OR for \textit{HsLCOM} is $1.92$, indicating that, as expected, a lack of cohesion increases the odds of inducing a refactoring operation.}

\revised{In conclusion, from our analysis it results that conceptual and readability metrics play a more important role in the model than structural metrics.  This finding is aligned with previous work aimed at applying conceptual metrics to suggest software refactoring \cite{DBLP:conf/icse/Bavota12} and modularization \cite{Bavota:emse2013}, and with findings of the seminal work about C3, indicating that such a metric is complementary to structural metrics \cite{DBLP:journals/tse/MarcusPF08}.}

\revised{None of the design flaws plays a statistically significant role. Although we expect that developers take care of removing smells, or try to ``make static analysis tools happy'', and although previous work has pointed the role of refactoring for improving code having a poor quality, \eg overly complex code \cite{Wang:icsm2009,Kim:fse2012,Silva:fse2016,Bavota:jss2015}, an evolutionary study on code smells indicate that smells mostly disappear when the source code is being rewritten, and only in less than 10\% of the cases because of a refactoring action \cite{DBLP:journals/tse/TufanoPBOPLP17}.}

	 %Differently, coping with  \textit{NPathComplexity} (number of acyclic execution paths through that method) is something more complicated, also considering that sometimes it is unavoidable that a piece of code is intrinsically complex. This could explain the fact that this design flaw is negatively correlated with refactoring.

Interestingly, process-related metrics are highly representative if compared to product-related metrics: five of the six considered process-related metrics are statistically significant. Previous bug fixes play a role: a unit increase of the \textit{FileRelatedToIssueFix} factor results in \revised{$2.09$} higher odds of applying a refactoring in the system. Not only classes subject to bug fixes are likely to be fault-prone in future~\cite{KimZWZ07,MoserPS08} but, since they are subject to (often quick-and-dirty) patches, they may necessitate refactoring actions. For related reasons, classes changing a lot (\textit{AvgLinesImpactedInCommit}) also need to be refactored, although the OR is smaller \revised{($1.93$)}.

\revised{Moving the attention to the metrics capturing the developers' experience, the \textit{Developer Class Experience} (\textit{LackFileExp}) has the highest OR. \revised{A unit increase of this factor related to the lack of specific experience (in terms of past commits) of developers that have recently modified a class, and therefore a decrease of experience increases the odds of refactoring by $8.93$ times.} In other words, changes applied by developers with little knowledge about a code component increase the need for restructuring it in the future. The general experience also plays statistically significant role, although the OR is relatively small (1.26).}

Finally, looking at the proximity to a release (\textit{DistanceNextRelease} and \textit{DistancePreviousRelease}), the mectrics indicate that refactoring operations are applied to the system far from a release of the system. \revised{More specifically, increasing the number of commits to a subsequent release, there are $1.43$ higher odds of applying a refactoring. However, results are not significant while looking at the number of commits from a previous release (\ie $\emph{p}-value = 0.11$).} Our findings confirm previous literature~\cite{kim:tse2014,Vassallo:scp2019} since developers are aware that some kind of refactorings may result  in the introduction of new faults~\cite{Bavota:scam2012} and in any case, refactoring represents a costly and risky operation~\cite{kim:tse2014}. For this reason, it is not very common to apply refactoring close to a new release of the software product. Furthermore, once released a new version of the software, developers likely tend to focus on bug-fixing activities instead of applying refactoring operations. 
In summary, based on our observations, refactorings are less likely to occur either immediately before major releases (developers focus on new features and, for what possible, on reliability of what they release), and immediately after (developers work on bug fixes). Instead, refactorings are more likely to happen in-between. Once again, note that this conclusion is based on purely observational data, and distance from the next release is unlikely to be used for prediction purposes (developers do not know when the next release is, unless a project or organization adopts very rigid release timelines).

We conclude RQ$_1$ stating that, on the one hand, by observing product metrics, only code readability plays a significant role. On the other hand, process-related metrics play a significant role. These are metrics related to previous changes and bug fixes, and to the experience of recent change authors.

\begin{figure*}[!h]
\begin{center}
\includegraphics[width=0.92\linewidth, angle =90, scale=1.3]{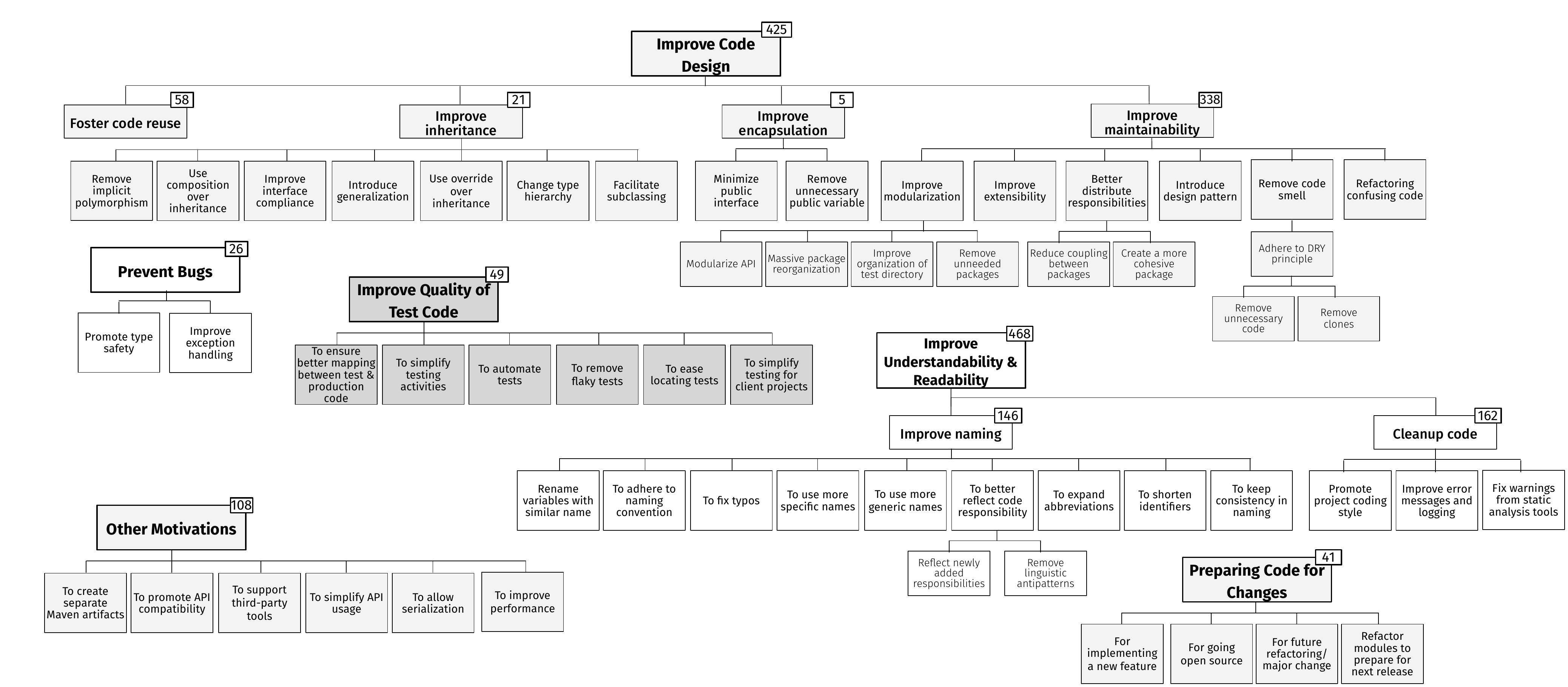}
\caption{Motivations behind refactoring operations}
\label{fig:motivations}
\end{center}
\end{figure*}

\begin{table}
	\caption{\revised{Statistics of refactoring operations labeled in the 551 analyzed PRs. The ``Freq.'' column reports the number of times that the annotators defined a tag explaining the rationale behind each specific type of refactoring operation. The total number of 1,117 tags is the result of the 1,223 tags we defined, excluding the 94 ``unclear'' (\ie cases in which the annotators did not manage to identify the rationale for the refactoring) and 12 ``false positives'' (\ie PRs that were unrelated to refactoring).}}
	\label{fig:operations_refactoring}
	
	\centering
	%\resizebox{\linewidth}{!}{
		\begin{tabular}{l | c | r r r | r r}
			\toprule
			\multirow{2}{*}{\textbf{Refactoring operation}} & \multirow{2}{*}{\textbf{Freq.}} & \multicolumn{3}{c |}{\textbf{When?}} & \multicolumn{2}{c}{\textbf{Tangled}} \\
			& & \textbf{Orig. intent} & \textbf{Collateral} & \textbf{After discuss.} & \textbf{Yes} & \textbf{No} \\ 
			\midrule
			Combination of refact. & 578 & 65\% & 3\% & 32\% & 86\% & 14\% \\ 
			Extract operation & 112 & 62\% & 3\% & 35\% & 62\% & 38\% \\ 
			Rename method & 69 & 47\% & 4\% & 49\% & 59\% & 41\% \\ 
			Rename class & 53 & 39\% & 6\% & 55\% & 56\% & 44\% \\ 
			Move class & 39 & 64\% & 8\% & 28\% & 49\% & 51\% \\ 
			Extract variable & 29 & 52\% & 35\% & 14\% & 72\% & 28\% \\ 
			Rename variable & 29 & 35\% & 14\% & 52\% & 83\% & 17\% \\ 
			Extract interface & 27 & 59\% & 11\% & 30\% & 59\% & 41\% \\
			Rename attribute & 22 & 35\% & 8\% & 57\% & 44\% & 56\% \\ 		
			Extract and move & 22 & 68\% & 0\% & 32\% & 55\% & 45\% \\ 
			Rename parameter & 19 & 69\% & 5\% & 26\% & 37\% & 63\% \\ 
			Move attribute & 15 & 53\% & 14\% & 33\% & 73\% & 27\% \\ 
			Move operation & 13 & 69\% & 0\% & 31\% & 39\% & 61\% \\ 
			Extract superclass & 10 & 70\% & 10\% & 20\% & 80\% & 20\% \\ 
			\midrule
			\textbf{Overall} & \textbf{1117} & \textbf{60\%} & \textbf{5\%} & \textbf{35\%} & \textbf{73\%} & \textbf{27\%} \\ 
			\bottomrule
			
		\end{tabular}
	%}
\end{table}

\subsection{\rqtwo}
\label{sub:qualitative}

Before digging into the results \tabref{fig:operations_refactoring} reports statistics about the refactoring operations we found in the analyzed PRs. Note that: (i) several refactorings can be applied in one PR, therefore the number of refactorings is higher than that of PRs; (ii) we only list refactoring operations we observed at least 10 times. However, the overall number of refactorings (1,117) also includes the instances related to refactoring types we do not show in \tabref{fig:operations_refactoring} (since having less than 9 occurrences). In most cases, developers do not discuss a specific refactoring operation. Instead, they rather provide a rationale for a combination of refactorings ($\sim$52\% of the cases). Then, \textit{extract operation} ($\sim$10\%) and renaming refactorings in general ($\sim$17\%, in total) are the ones more discussed by developers. Surprisingly, we found that only a small percentage (5\%) of refactorings were done collaterally, \ie without mentioning them at all. Instead, many of them were done as the original intent of the PR ($\sim$60\%) or after discussing with other developers ($\sim$35\%). 

Some refactoring operations, such as \textit{extract variable} and \textit{rename variable}, were performed collaterally more often, given their ``local'' nature: a variable name only matters in the methods in which it is declared, while a class name can possibly impact the whole system. It is worth noting that developers perform most of the renaming operations after they receive feedback from their peers. This shows that names are often discussed in PR reviewing activities. Finally, in line with Murphy-Hill \etal~\cite{Emerson:tse2011}, we found that about a fourth of the refactorings are tangled with other changes.

\revised{\figref{fig:motivations} depicts the taxonomy of refactoring motivations we have identified. It comprises six root categories: (i) \emph{Improve Code Design} groups refactoring operations targeting an improvement of the system design, \eg by fostering the reusability of code; (ii) \emph{Improve Understandability \& Readability} includes refactorings aimed at reducing the effort to read and understand code, \eg by renaming identifiers; (iii) \emph{Improve Quality of Test Code} groups all refactorings performed to improve the quality of the test code or ease the testing process; (iv) \emph{Prevent Bugs} identifies refactorings performed to prevent the future introduction of bugs; (v) \emph{Preparing Code for Changes} includes refactorings performed in preparation of other changes, \eg refactoring the code before implementing a new feature; (vi) finally, \emph{Other Motivations} groups those motivations that cannot be classified into one of the previous categories.} 

It is important to point out that some of the categories of the taxonomy are not mutually exclusive. For example, a refactoring aimed at improving code  readability, is also likely to improve maintainability. However, readability can be improved for multiple purposes (\eg simplify testing), and, for this reason, we separated these categories. We acknowledge that other choices in terms of categories and assignment of instances to these categories are possible. \rev{Also, the hierarchical organization of the categories only indicates that child categories are specialization of their parent categories, while it does not imply that two categories at the same hierarchy level represent motivations at the same level of abstraction (\eg \emph{Improve Inheritance} and \emph{Foster Code Reuse} appear at the same level, but the former is a more concrete motivation as compared to the latter).}

\figref{fig:motivations} also reports for each category of ``motivations'' the number of PRs in which we found related refactoring operations. For readability purposes, we only report these numbers for the main categories. Note that the number for a parent category does not correspond to the sum of the children, because some PRs were only assigned to the parent category, as the motivation was not specific enough.
%Also, it is worth noticing that the number of refactorings in a parent category (\eg \emph{Understandability \& Readability}) might not correspond to the sum of the refactorings present in its child categories (\eg \emph{Refactoring confusing code}, \emph{Renaming}, and \emph{Cleanup Code}). This is due to the fact that during our labeling process, we tried to assign in each case the most specific label (\ie motivation for refactoring) we could infer by looking at the code changes and at the discussion in the PR. Sometimes we were able to assign very specific motivations (\eg \emph{Rename variables with similar name}), in other cases developers only documented the general motivation. Thus, while in the former case the refactorings in the PR contribute to the \emph{Rename variables with similar name} category, to its parent class \emph{Refactoring confusing code}, and to the root node \emph{Understandability \& Readability}, in the latter case it only contributes to the root node. This means that the number of refactorings in the parent notes is always greater than or equal to the ones in the children nodes.
Also, the sum of refactoring instances in all root nodes does not correspond to the total number of \manually manually analyzed PRs because some PRs comprise refactorings falling into multiple categories, and we labeled some refactorings as \emph{Unclear} (94) and discarded 12 PRs as \emph{False Positive}. 

We compared our taxonomy with the list of 44 motivations derived by Silva \etal~\cite{Silva:fse2016} for 12 frequently applied refactoring operations (see Tables 3 and 4 in~\cite{Silva:fse2016}). In particular, two of the authors tried to map Silva \etal's motivations into our taxonomy, to see whether they were covered or not. Note that the mapping is not one-to-one since one motivation identified by Silva \etal~\cite{Silva:fse2016} may be mapped to more than one category in our taxonomy, as well as one of our categories can group more than one of their motivations. This is expected since their motivations and the categories in our taxonomy have been derived by using two different methodologies. Indeed, while we have categorized the possible motivations behind the application of refactoring operations by looking at the discussions in PRs, Silva \etal~\cite{Silva:fse2016} have asked the reasons behind specific instances of refactoring operations to the original developer who has applied it.

\revised{Only 3 out of the 44 motivations from Silva \etal~\cite{Silva:fse2016} cannot be mapped in our taxonomy}. The main reason is that for these three instances (\ie \emph{Enable recursion}, \emph{Convert to top-level container}, and \emph{Convert to inner class}) it was unclear to us the actual motivation behind the refactoring. For example, enabling recursion could be done to improve performance as well as to improve code readability. However, this high overlap between the two sets of motivations (i) validates \rev{and generalizes} the work done by Silva \etal and (ii) supports the comprehensiveness of our taxonomy. \revised{As reported in \tabref{fig:comparison}, our taxonomy features 16 inner categories that are not covered in~\cite{Silva:fse2016} (\eg \emph{To Ensure Better Mapping Between Test And Production Code}), 3 inner categories that are only partially covered (\eg \emph{Preparing Code for Changes}), and 6 inner categories (\eg \emph{Forster Code Reuse}) that are completely covered.} We provide in our replication package~\cite{replication} a spreadsheet reporting the mapping between the two taxonomies.

In the following, we discuss each root category, reporting interesting examples and outlining implications for researchers and practitioners (indicated with the \faLightbulbO~icon), as well as highlighting the differences with the taxonomy provided by Silva \etal~\cite{Silva:fse2016}. The complete list of manually analyzed PRs together with their refactorings/assigned tags is publicly available~\cite{replication}.

\begin{table}
	\caption{\revised{Comparison with refactoring motivations found by Silva \etal~\cite{Silva:fse2016}: $\uparrow$ $\uparrow$ highlights a perfect match (the motivation also emerges from their study); $\uparrow$ highlights a partial match (we found additional, specific motivations); $\downarrow$ $\downarrow$ stands for a mismatch (the motivation was not found in their study).}}
	\label{fig:comparison}
	\footnotesize
	\centering
	\resizebox{\linewidth}{!}{
		\begin{tabular}{l | l | c}
		\hline
		\textbf{Root Category} & \textbf{Inner Category} & \textbf{Match}\\
		\hline
		\multirow{4}{*}{Improve Code Design} 
		& Foster code reuse & $\uparrow$ $\uparrow$\\
		%& Adhere to Dry principle & $\uparrow$ $\uparrow$\\
		& Improve inheritance & $\uparrow$ \\
		%& Remove code smells &$\uparrow$ $\uparrow$\\
		& Improve encapsulation & $\downarrow$ $\downarrow$\\
		& Improve maintainability & $\uparrow$ $\uparrow$\\
		\hline
		\multirow{2}{*}{Improve Understandability \& Readability} 
		& Improve naming & $\uparrow$\\
		& Cleanup code &$\downarrow$ $\downarrow$\\
		\hline
		\multirow{6}{*}{Improve Quality of Test Code} 
		& To ensure better mapping between test and production code & $\downarrow$ $\downarrow$\\
		& To simplify testing activities & $\uparrow$ $\uparrow$\\
		& To automate tests & $\downarrow$ $\downarrow$\\
		& To remove flaky tests & $\downarrow$ $\downarrow$\\
		& To ease locating tests & $\downarrow$ $\downarrow$\\
		& To simplify testing for client projects & $\downarrow$ $\downarrow$\\
		\hline
		\multirow{5}{*}{Preparing Code for Changes} 
		& For implementing a new feature & $\uparrow$  $\uparrow$\\
		& For going open source & $\downarrow$ $\downarrow$\\
		& For future refactoring/major change & $\downarrow$ $\downarrow$\\
		& Refactor modules to prepare for next release & $\downarrow$ $\downarrow$\\
		\hline
		\multirow{2}{*}{Prevent Bugs} 
	& Promote type safety & $\downarrow$ $\downarrow$\\
	& Improve exception handling & $\downarrow$ $\downarrow$\\
	\hline
	\multirow{6}{*}{Other Motivations} 
	& To create separate Maven artifacts & $\downarrow$ $\downarrow$\\
	& To promote API compatibility & $\uparrow$ $\uparrow$\\
	& To support third-party tools & $\downarrow$ $\downarrow$\\
	& To simplify API usage & $\downarrow$ $\downarrow$\\
	& To allow serialization & $\downarrow$ $\downarrow$\\
	& To improve performance & $\uparrow$ $\uparrow$\\
	\hline
		\end{tabular}
	}
\end{table}

\textbf{Improve Code Design (425 instances).} Unsurprisingly, a large proportion of the analyzed refactorings are aimed at improving code design from several perspectives \cite{fowler-refactoring}. In 58 of these, the refactorings are aimed at making source code easier to be reused (see \emph{Foster code reuse} in \figref{fig:motivations}). In 42\% of these cases, this was accomplished through a combination of several refactoring operations, while in the remaining 58\% specific refactorings were applied in isolation. When this happened, almost always (91\%) an operation aimed at extracting a code component from an existing one was applied. In particular, \emph{extract method} operations were performed in 70\% of cases, to extract a small piece of functionality from an existing method thus avoiding code duplications and fostering the reuse of the extracted code. For example, during the code review of the PR \#626 in the \texttt{nakadi} project~\cite{nakadi626}, the reviewer observed that two of the implemented methods were ``\emph{almost the same except the very last line}'' and suggested to ``\emph{extract a helper method}'' in such a way to reduce code duplication and also allow other methods, in future, to reuse the same functionality. This was accomplished through an extract method refactoring. In other cases, the refactoring was more substantial and directly justified by the need for reusing specific pieces of functionality, as discussed in the PR \#488 of the \texttt{dropwizard} project~\cite{dropwizard488}.
%\footnote{\url{https://github.com/dropwizard/dropwizard/pull/488}}. 
Here the contributor explains, when submitting the PR: ``\emph{I was looking at starting/stopping a Dropwizard app in Cucumber tests and DropwizardAppRule has all the functionality I need but obviously it doesn't expose startIfRequired and stop methods. I'd happy to extract a DropWizardAppTestSupport class from DropwizardAppRule}''. After approval, this triggered an \emph{extract class} refactoring. This last example is interesting for several reasons. \faLightbulbO~First, extract class is a non-trivial refactoring possibly having substantial ripple effects in the system, with the obvious possibility of introducing bugs. For example, the discussed commit impacted a total of 499 lines of code, thus showing that code reuse is a strong motivation for triggering refactoring operations. Second, while many approaches to identify extract class refactoring opportunities have been proposed (see \eg~\cite{FokaefsTSC11,Bavota:emse2014b}), they focus on the identification of complex classes implementing several responsibilities (\ie \emph{God} or \emph{Blob} classes~\cite{Brown98-AntiPatterns}) that could be split into several classes. \revised{The class subject of the \emph{extract class} refactoring (\ie {\tt DropwizardAppRule}) is a fairly simple class composed by 156 effective LOC (excluding comments and blank lines) that is unlikely to be reported by refactoring recommenders as a candidate for \emph{extract class} refactoring. To cope with these cases, these recommenders could be combined with clone detectors~\cite{Roy:2018} to factor out a class to be used by multiple other ones. Note that these ``special'' cases should complement the more standard refactoring recommendations done for complex and low-cohesive classes. Indeed, as shown in our RQ$_1$, classes characterized by a low cohesion as assessed by the C3 and HsLCOM metrics are more likely to be subject to refactoring operations.}

Many (338) of the refactorings performed in the code design taxonomy aimed at \emph{Improve Maintainability} (see \figref{fig:motivations}). In this category, refactorings aimed at improving the modularization were often implemented through simple move class refactorings, while we rarely observed massive package reorganizations (7 cases). \faLightbulbO~This is in line with recommendations from previous literature, suggesting that approaches performing big-bang remodularization through clustering algorithms have limited applicability, and techniques suggesting fine-grained and incremental adjustments to software modularization should be preferred~\cite{Hall:icsm2012,paixao:tec2018}. Also, it was interesting why developers decided to perform remodularization.  
For example, in some cases move class refactorings are performed to group, in specific ``API-related'' packages, utility classes potentially useful in different parts of the system and/or to third-party components, \eg PR \#324 from \texttt{DSpace} ``\emph{I suggest to move this class in dspace-api as it will be useful to port this feature to JSP UI as well}''~\cite{DSpace324}. While these changes might look suboptimal from the cohesion-coupling point of view (\ie they could generate a low-cohesive package) they are justified by a clear rationale. \faLightbulbO~As also observed for the approaches automating \emph{extract class} refactoring, tools recommending modularization solutions (see \eg ~\cite{Anquetil:wcre1999,Kuhn:ist2007,Maqbool:tse2007,PraditwongHY11}) just strive to maximize the cohesion-coupling trade-off. Given the availability of historical data, they could also learn from previous changes what a meaningful modularization is from the developers' perspective. While learning code changes is already an active research field~\cite{Tufano:icse2019}, no previous work has attempted to design refactoring recommenders learning from developers' activities what a meaningful refactoring is in a given context.

Removal of code clones is one of the two motivations behind the refactorings in the \emph{Adhere to DRY principle} (Don't Repeat Yourself) subcategory (child of \emph{Remove Code Smell}), together with the removal of unnecessary code. Note that this category is strictly related to the \emph{Foster code reuse} one. Indeed, some of the analyzed PRs fall into both these categories because factoring out duplicates also creates a more generic code element (\eg a class or method) that can be further reused. For example, PR \#366 in the \texttt{fineract} project~\cite{fineract366}
%\footnote{\url{https://github.com/apache/fineract/pull/366}} 
can be seen as an example of improving reusability by adhering to the DRY principle, since it features an \emph{extract method} refactoring suggested by the reviewer and avoiding code duplication while allowing the reuse of a piece of functionality now embedded in the extracted method. \faLightbulbO~This confirms once more the relevance for practitioners of clone detectors~\cite{Roy:2018} as well as of refactoring tools aimed at removing clones~\cite{Tsantalis:2017} and encourages their use in the Continuous Integration (CI) pipeline, as advocated by Duvall \etal~\cite{Duvall:2007:CII:1208841}. 
%Indeed, we found many cases the clone removal opportunity  could have  been automatically identified, and possibly fixed, by integrating clone detectors, and recommenders to suggest their removal, in the CI process.

Concerning the removal of unnecessary code, besides cases simply related to removing unused \texttt{imports}, we found refactorings performed to remove redundant code (\eg PR \#1481 from \texttt{testng}~\cite{testng1481}.
%\footnote{\url{https://github.com/cbeust/testng/pull/1481}}). 
\faLightbulbO~While some work has investigated the automatic identification of redundant code in software systems~\cite{Leit2004,Kawrykow:ase2009}, the provided support is still very limited to specific redundancy cases (\eg those related to API usages~\cite{Kawrykow:ase2009}) or programming languages (\eg LISP~\cite{Leit2004}). The 55 cases related to refactorings motivated by the removal of unnecessary code suggest room for more research in this field.

Concerning the operations targeting a better distribution of the responsibilities across code components, one very interesting example comes from the \texttt{DSpace} project (PR \#1083~\cite{DSpace1083}).
%\footnote{\url{https://github.com/DSpace/DSpace/pull/1083}}). 
This PR implements a massive refactoring aimed at ensuring a better ``separation of concerns/responsibilities'' for an API module, and has been subject to votes by the community, because the refactored API was not backward compatible. Despite this issue, the merging has been approved thanks to the numerous positive advantages brought by the refactored API: ``\emph{makes it much easier to achieve future goals on our Roadmap, especially, moving us towards potentially better support of third-party modules}'', ``\emph{it cleans up one of the messiest areas of our existing API [\dots]}''.~\faLightbulbO~This case shows the non-trivial trade-offs that developers should consider in case of massive refactoring: for example, smartphones have limited battery life and they require software optimized to reduce the energy consumption. State-of-the-art refactoring recommenders~\cite{Tsantalis:tse2009,Bavota:emse2014b} ignore the heterogeneity of modern software, and the different priorities that non-functional requirements, possibly more important (\eg maintainability, performance, backward API compatibility) may have in different contexts. Future work should consider integrating into these recommender systems the possibility to define a priority list of non-functional properties that developers are or are not willing to sacrifice when applying refactoring. This would allow generating more meaningful and sensible refactoring recommendations.

\revised{The two sub-categories described above (\emph{Foster code reuse} and \emph{Improve maintainability}), \ie the ones highly represented in our study, have a complete matching with the motivations identified by Silva \etal~\cite{Silva:fse2016}}. This confirms their findings, and stresses once more the importance from the developers' perspective of improving both the reusability and maintainability of code, especially when discussing whether to accept or not a PR.

Other less represented subcategories in the \emph{Improve code design} taxonomy include refactorings aimed at improving the usage of inheritance (21 instances) and the ones working on the encapsulation (5). In these cases, considering the low number of instances belonging to each category, it is quite obvious that while comparing with the motivations by Silva \etal we found that these reasons did not emerge from their study. The only exception is the one related to inheritance that is only partially covered, as shown in \tabref{fig:comparison}.
 %OK

\textbf{Improve Understandability \& Readability (468 instances).} The majority of refactorings we found in the manually-analyzed PRs aim at improving understandability and readability of source code. This supports the findings of RQ$_1$, which indicate a significant correlation (and high OR) of readability metrics with refactoring operations. In this category, 146 refactorings were done to improve naming. The observed renamings had a variety of motivations (see \figref{fig:motivations}), ranging from fixing typos to keeping naming consistency throughout the project. Naming decisions were often carefully discussed, showing their importance for developers. An interesting example of discussion about naming is the PR \#150 of the \texttt{optaplanner} project~\cite{optaplanner150}.
%\footnote{\url{https://github.com/kiegroup/optaplanner/pull/150}}. 
The original intent of the PR was to add a new feature, but the author explicitly asked for feedback about the naming of a new interface he extracted: ``\textit{we should discuss the naming and the usage of the Solver\-Problem\-Benchmark\-Result interface}''. Such a name was changed after the discussion: ``\textit{renamed to Benchmark\-Result as agreed in a meeting''}. In the same PR the developers also discussed several other names in the contributed code. For example, the PR author introduced a boolean field named \texttt{has\-Non\-Default\-Sub\-Single\-Count}; another developer asked why such a value was introduced, since the name was not clear enough. The discussion triggered not only a renaming operation, but also a type change (from boolean to integer) to represent additional information that could be useful in future: ``\textit{\texttt{maximum\-Sub\-Single\-Count} is the best name here, as it gives us more potential information for the future at no cost}''. 

We also found cases of renaming aimed at better reflecting the code responsibility (38). A representative example is in PR \#251 of the \texttt{kafka-connect-elasticsearch} project~\cite{kafka251},
%\footnote{\url{https://github.com/confluentinc/kafka-connect-elasticsearch/pull/251}},
 in which one of the reviewers suggested to rename a test method to something very specific and clearly depicting the responsibility of the test case: ``\textit{you could change the name of the test method to something like test\-Create\-And\-Write\-To\-Index\-For\-Topic\-With\-Uppercase\-Characters. I like test names that read like the condition they are testing}''. Other cases aimed at removing linguistic antipatterns defined in the literature by Arnaoudova \etal~\cite{ArnaoudovaPAG13} and known to have negative effects on the understandability of code~\cite{Fakhoury:icpc2018}. \faLightbulbO~This is only one of the studies linking the poor quality of identifiers to difficulties experienced by developers in code comprehension~\cite{Takang:jpl1996, Lawrie:isse2007, Lawrie:icpc2006naming, deibenbock2005, LawrieFB06}. Our findings show that developers care about the quality of identifiers and carefully discuss their choice. 
 
\faLightbulbO~Most of the rename refactoring recommendation approaches aim at fostering the usage of consistent naming~\cite{Alla2014a, lin2017investigating}, while only a single attempt has been done, to the best of our knowledge, to recommend rename method refactorings with the goal of better reflecting the responsibilities implemented by the code~\cite{Alon:2019}. Our results show that more effort in this direction is needed since this is the scenario in which developers more frequently perform rename refactorings. Also, \faLightbulbO~ the high number of rename refactorings implemented as a consequence of code review indicates the possibility to mine this data to evaluate automated rename refactoring techniques~\cite{Alla2014a, lin2017investigating}: the originally submitted identifier represents an opportunity for rename refactoring while the one adopted after the code review process can be used as reference of a good refactoring. This would avoid the evaluation of the rename refactoring techniques in artificial scenarios.

Looking at \tabref{fig:comparison}, and considering that developers can modify the names of packages, classes, variables or methods for different reasons, we can state that our \emph{Improve Naming} category is only partially covered by the motivations reported in the previous study by Silva \etal~\cite{Silva:fse2016}. For instance, while both studies identify the need for adhering to naming conventions, for keeping consistency in naming, or for better representing code responsibilities, in our taxonomy we also found cases where the renaming occurs to fix typos, shorten identifiers and expand abbreviations.\revised{~\faLightbulbO~The latter challenge (\ie expansion of abbreviations in code identifiers) has been vastly investigated in the software engineering research literature (see, for example, the works by Lawrie \etal \cite{Lawrie:ICSM2011}), with approaches proposed and empirically evaluated. However, to the best of our knowledge, there are no ready-to-use tools that, for example, can be integrated in a CI pipeline and can recommend to developers identifiers to expand at commit time. The implementation of such a tool is a clear next step to perform in this research field.}

We also found several refactorings implemented to make the source code less confusing. Such changes involved improvements to both names and structural aspects. For example, we found an interesting example in PR \#599 of the \texttt{htsjdk} project~\cite{htsjdk599}. \revised{Note that the refactoring performed in this PR is an \emph{extract interface} rather than a rename, that resulted in the interface \texttt{CRAM\-Reference\-Source} implemented by the class \texttt{Reference\-Source}. The main goal of the refactoring was to improve code reusability: for this reason, we include such a case in our taxonomy under the \emph{Foster code reuse} category. However, it is interesting to discuss this refactoring in the context of renaming: during the code review process, 
%\footnote{\url{https://github.com/samtools/htsjdk/pull/599}}. 
 one of the reviewers argued that the chosen names were confusing because he expected an inheritance relationship in the opposite direction (\ie \texttt{CRAM\-Reference\-Source} implements \texttt{Reference\-Source}) by reading the names alone}. As a consequence, \texttt{Reference\-Source} was renamed to \texttt{CRAM\-ReferenceSourceImpl}, making the relationship between the two classes more evident. \revised{~\faLightbulbO~This naming issue could be characterized as a sort of linguistic antipattern, and shows that the original catalog of these antipatterns defined by Arnaoudova \etal~\cite{ArnaoudovaPAG13} could be expanded by analyzing recommendations provided by reviewers in a code review process}. 
%The reasons why poor naming is introduced can be many-fold, and, reflecting from the quantitative analysis of RQ$_1$, these could include changes by developers with limited or no past change history on the specific class, as well as (urgent) bug fixes. \MAX{it would be great if there is some evidence of that...}

Finally, we found many cases in which the developers made more generic clean-ups in the code (162 instances), to improve the coding style and, in some cases, the quality of error messages and logging. Some of these changes were performed as a result of tools' recommendations. For example, the PR \#346 of the \texttt{spring-amqp} project~\cite{amqp346} 
%\footnote{\url{https://github.com/spring-projects/spring-amqp/pull/346}} 
fixes warnings raised by SonarQube. \revised{~\faLightbulbO~This suggests that developers are willing to fix issues identified by automatic tools}. However, in our sample of PRs, SonarQube was the only tool mentioned in many discussions. \revised{Note that a specific analysis of the extent to which static analysis tool warnings are removed was not in scope of our work (but rather addressed in related literature \cite{Couto2011StaticCA,KimE07,spacco2006tracking}); this is the reason why, in RQ$_1$, we only considered two tools --- DECOR and PMD --- that could raise warnings that triggered refactoring operations}.

More than 60\% of the clean-up operations were done as part of the original intent of the PRs. Differently from other categories, we found very little discussion among developers regarding clean-ups. This is likely due to (i) the limited impact that these clean-ups generally have; and (ii) a general agreement on the need for improving code quality.
 %OK

\revised{\textbf{Prevent Bugs (26 instances).} These refactorings are motivated by the will to prevent bugs, for example through a better exception handling or by promoting type safety. Note that these are changes that preserve the program's behavior. For this reason, we contemplated them in our taxonomy, even if they do not belong to the canonical cases of refactorings, such as those defined by Fowler~\cite{fowler-refactoring}). This is why those categories are completely uncovered in the 44 motivations provided by Silva \etal~\cite{Silva:fse2016}. Indeed, they studied the reasons behind specific refactoring operations that are detected by \reftool and inline with those defined by Fowler~\cite{fowler-refactoring}.}

\revised{Some PRs are explicitly motivated by the will of improving the exception handling mechanism. This is the case for PR\#933 of the {\tt nakadi} project \cite{nakadi933}, in which the developer implements several different refactorings (\eg \emph{rename class}, \emph{move class}) to improve the overall handling of the exceptions in the project. Other PRs, instead, simplify the handling of exceptional conditions. For example, in PR\#1067 from the {\tt htsjdk} project \cite{htsjdk1067}, the developer fixes a possible issue caused by the invocation of the method {\tt mFile.\-getSource()} within several exception messages. Indeed, in specific cases the {\tt mFile} object could be {\tt null}, leading to the throwing of a {\tt NullPointerException}. For this reason, the developer implemented an \emph{extract method} refactoring, creating the method {\tt getSource()} which returns the value of {\tt mFile.\-getSource()} when {\tt mFile} is not {\tt null}, and a constant string otherwise. This allowed to easily prevent {\tt NullPointerException} by replacing the many usages of {\tt mFile.\-getSource()} with an invocation to the newly created {\tt getSource()} method. \faLightbulbO~This is an interesting application of \emph{extract method} refactoring, since it aims at refactoring a very small clone, \ie a method invocation reused, in the same way, in different parts of the code. Extract method is widely applied in the refactoring of clones \cite{Krishnan:2014}. However, the focus is usually on more complex clones sharing several statements, rather than on the identification of refactoring opportunities that, as in the discussed case, involve few code tokens but can have a positive impact on the reliability and maintainability of the system.}

\revised{Another very interesting example is the PR\#238 from the {\tt minio-java} project \cite{minio238}. Here the developer replaced general, unchecked exceptions such as {\tt NullPointerException}, with more specific and checked ones. With ``unchecked'' we refer to those exceptions that in Java can be thrown without declaring them in the method signature. For example, in the specific case of PR\#238, the method {\tt getClient} of the class {\tt Client} was throwing a {\tt NullPointerException} in specific situations: note that this was an intended behavior of the method, \ie there was an explicit {\tt throw new NullPointerException()} in the code. However, in the method signature the only visible exception was {\tt MalformedURLException}. Through the implemented changes, including \emph{class rename} refactoring, the more general exceptions have been specialized (\eg to {\tt ClientException} in the case of the {\tt getClient} method), forcing the exposure in the signature of the thrown exception. This has the double effect of (i) giving developers compilation errors if they do not catch the thrown exception, thus preventing bugs; and (ii) using more expressive exception names. \faLightbulbO~Note that the usage of unchecked exceptions in Java code should not be considered as a ``bad smell'' since, in general, unchecked exceptions should be used to reveal bugs, while checked exceptions to throw errors that the program should handle \cite{Chen:jss2009}. However, the misuse of unchecked exceptions where checked ones are needed can lead to higher chances of introducing bugs (as in the case of PR\#238). The automatic identification of these situations is, to the best of our knowledge, a problem still not faced in the research literature.}

%\revised{Finally, in PR\#282 of the {\tt clickhouse-jdbc} project \cite{clickhouse282}, a rename refactoring aimed at preventing bugs was performed as result of the code review process. We classified this instance in the \emph{Prevent bugs} root category, since the reviewer, commented on the new method }

 %OK

\revised{\textbf{Preparing Code for Changes (41 instances).} This category includes refactorings facilitating the implementation of new features or of other planned activities. Refactorings preparing for future changes are usually implemented in dedicated PRs including major refactorings.}
%\footnote{\url{https://github.com/zendesk/maxwell/pull/392}}). 

Looking at the comparison highlighted in \tabref{fig:comparison}, it is possible to state that our taxonomy provides more insights compared to the list of motivations in the previous study. Indeed, the category \emph{Preparing Code For Changes} contains some motivations already highlighted in \cite{Silva:fse2016} while others missed such as the need for refactoring operations aimed at moving to open-source. \revised{An interesting case in this category is represented by PR \#317 of the \texttt{sagan} project~\cite{sagan317}. 
%\footnote{\url{https://github.com/spring-io/sagan/pull/317}}. 
As mentioned in the title, the goal of the PR is to \emph{``refactor in preparation for open source''}. Note that this is kind of an exception in our taxonomy, and we decided to put it into the \emph{Preparing Code for Changes} category just because open sourcing a project is, in some way, a decision taken to foster the future evolution of the project. Code refactoring is one of the action items in a checklist defined in issue \#179 for preparing the project to be open-sourced~\cite{sagan179}; other action items included, for example, the introduction of installation and configuration instructions. \faLightbulbO~This suggests that an appropriate code cleanup, including refactoring, should be part of packaging checklists when putting a project in the open-source.}
 
\revised{Another example of refactoring performed to accommodate other changes is in PR\#136 of the {\tt zhcet-web} project \cite{zhcet136}, where the developer extracted the class {\tt CryptoUtils} from {\tt SecurityUtils} to accommodate the implementation of new functionalities (\eg the {\tt decrypt} method) in a suitable class (\ie {\tt CryptoUtils}).}	

\faLightbulbO~Tools supporting preemptive refactoring are lacking in the literature. Indeed, the only effort in this direction is the work by Pantiuchina \etal~\cite{Pantiuchina:icpc2018} in which, however, the focus is on identifying classes that will be affected by code smells in the future, thus recommending them for a preemptive refactoring action. \faLightbulbO~Our manual analysis indicates that a novel family of recommender systems able to suggest developers how to refactor the code in order to ``accommodate'' the implementation of a given change request could be valuable.
 %OK

\textbf{Improve Quality of Test Code (49 instances).} As the production code might be in need for refactoring, this also holds for test code~\cite{van2001refactoring}. The quality of the test code is also assessed in the context of PR discussion and code reviews, as observed by Spadini \etal~\cite{SpadiniPBHBB19}. We found 49 test code refactoring cases, 37\% of which performed with a specific type of refactoring operation, and 63\% with multiple operations. The observed refactorings include changes similar to those performed on production code, \eg better distribution of responsibilities to have a better mapping between test and production code, see \eg PR \#1071 in the \texttt{error-prone} project~\cite{error1071}
%\footnote{\url{https://github.com/google/error-prone/pull/1071}}
 in which the author comments ``\emph{[...] separate the tests into logical classes}''.

Other cases we found concern the removal of flaky tests which introduce non-determinism in test outcome~\cite{micco2016flaky}. In the \texttt{microprofile\--fault\--tolerance} project, 
%\footnote{\url{https://github.com/eclipse/microprofile-fault-tolerance/pull/363}}, 
the PR \#363 \cite{mft363} aims at removing flaky test: ``\emph{The tests test\-Circuit\-Initial\-Success\-Default\-Success\-Threshold and test\-Circuit\-Late\-Success\-Default\-Success\-Threshold were moved to an independent test to avoid dependencies between tests that use the same bean [...] that can generate possible failures when the circuit breaker leaves open [...]}''.

More interesting and specific for test code are the refactorings performed to improve testability. In PR \#73 of the \texttt{WPS} project~\cite{wps73}
%\footnote{\url{https://github.com/52North/WPS/pull/73}}
a developer performed an \emph{extract class} refactoring in production code motivated by the will of simplifying integration testing: ``\emph{The functionality to create a GT\-Vector\-Databinding out of shapefiles was removed from the Generic\-File\-Data class and moved to a new Generic\-File\-Data\-With\-GT class. Due to this change, the processes used for the integration tests do not depend on Geo\-Tools anymore [...] also the tests now use only local resources}''. \faLightbulbO~This example shows how refactoring performed on production code can have an impact on many software quality aspects (in this case, testability). Cases like this one suggest to always ponder the positive and negative impacts of refactoring beyond maintainability \eg some refactoring actions can aid testability, while others might improve  maintainability at the cost of testability. \revised{\faLightbulbO~Clearly, considering this aspect in the context of a refactoring recommender is far from trivial, given the need for automatically assess the testability of a given component.}
Besides, the presence of refactorings specifically aimed at improving testability shows room for approaches aimed at recommending such kind of operations. 

While the motivations of refactorings for simplifying testing activities were already presented in the work by Silva \etal~\cite{Silva:fse2016}, our taxonomy provides other new categories such as those aimed at removing flakiness, at automating testing activities, or at improving the overall testability of the project under development.

%While in this PR the improvement of testability was the refactoring goal, in other cases, as previously said, the impact of refactorings on a number of different quality aspects should be carefully considered before changing the code (\ie an improvement in maintainability can have a cost in terms of testability).

 %OK

\revised{\textbf{Other Motivations (108 instances).}
In this category, we put all motivations that did not find their place among other root categories.}

\revised{Thirty-seven PRs were performed to improve software performance. This is not surprising and in line with Fowler \cite{fowler-refactoring}, who observed that the internal program structure is closely related to its performance due to better optimization opportunities. Moreover, the latter also emerges from the motivation provided in the previous study by Silva \etal~\cite{Silva:fse2016}.
Also, our quantitative results of RQ$_1$ indicate that refactorings have a high chance to occur on classes frequently subject to bug fixes, which may have affected performance, especially in the case of quick patches. A concrete example is the PR \#1577 of \texttt{AmazeFileManage}~\cite{amaze1577} Android application. 
%\footnote{\url{https://github.com/TeamAmaze/AmazeFileManager/pull/1577}}, an open source Android app. 
In a linked issue, a user reports that when she is ``\emph{copying a large file using SFTP, the process can take more than 1 minute, so the phone goes in stand by mode}''. PR \#1577 improves the I/O performance of this feature through refactoring.}

\revised{\faLightbulbO~This example confirms the importance of specific non-functional attributes (in this case, performance) for different types of software (in this case, a mobile app). Also, once again, it points to the need for developing refactoring techniques able to consider this heterogeneity of non-functional requirements rather than mainly focusing on maintainability as done in state-of-the-art refactoring tools \cite{Tsantalis:tse2009,Bavota:emse2014b}. Indeed, to the best of our knowledge, only a few authors have developed refactoring techniques having the improvement of performance as the main objective~\cite{Dig:Software,Zhang:2012,Arcelli:qosa2012}; however, these approaches either target very specific performance issues~\cite{Dig:Software,Zhang:2012} or are designed to work on models rather than on source code~\cite{Arcelli:qosa2012}.}

\section{Threats to Validity} \label{sec:threats}
%%%%%%%%%%%%%%%%%%%%%%
%%%%%%%%%%%%%%%%%%%%%%

\textbf{Construct validity.} A source of inaccuracy is represented by the automated refactoring detection. However, \reftool has been reported to exhibit a very high precision (98\%) and recall (87\%)~\cite{Tsantalis:icse2018}. This threat is mitigated at least in RQ$_2$, where refactorings have been manually reviewed.
To identify bug fixes, we used an approach matching regular expressions onto commit messages~\cite{DBLP:conf/icsm/FischerPG03}, as also done in previous work \cite{RayPFD14}. To limit threats due to this heuristic~\cite{AntoniolAPKG08}, two authors independently analyzed, for each project we considered, a random sample of commits classified as a bug fix to mark true and false positives. After discussing disagreements, only 8\% of the analyzed commits resulted to be false positive bug fixes (mostly related to CheckStyle fixes).

In RQ$_1$, we only analyzed the correlation between the presence of any refactoring with various metrics. While it may be interesting correlating specific types of refactorings with metrics, our qualitative analysis showed that refactoring goals are often achieved through a combination of refactorings. \rev{To build the explanatory model of RQ$_1$, we have selected a broad set of metrics capturing different aspects of software product and process. It is important to note that the aim was to correlate such metrics with the presence of at least one refactoring action of any kind. Building models for specific refactoring types is out of scope of this paper and could, possibly, require to identify further specific indicators.}

\rev{In RQ$_2$ we identified refactoring-related PRs as those having (i) one of their commits containing a refactoring identified by \reftool, and (ii) a refactoring-related keyword in their title. Such selection criteria can result in false negatives (\ie missing some refactoring-related PRs) and, in turn, this may have resulted in missing categories in our taxonomy. Indeed, it is possible that our taxonomy is only representative of the motivations behind PRs that can be captured through the adopted selection criteria.}

\rev{As context for our study we targeted non-personal/toy projects having a substantial change history to study and being active. For this reason, we defined a number of selection criteria (\ie at least 5 contributors, 1 fork, 500 commits, 100 PRs, and one recent commit) that, however, may fail in capturing the type of systems we were interested in.}
 
\textbf{Internal validity.} In the quantitative analysis (RQ$_1$), although we tried to capture factors from different dimensions (\ie different kinds product and process metrics), there could be many other factors that could have influenced the need for refactoring. We mitigated this threat through (i) the use of a mixed-model considering project as random effect, and (ii) the qualitative analysis of RQ$_2$. \revised{It is important to note that the aim of RQ$_1$ is to mainly identify correlations between metrics and refactoring activities, and not about claiming any causation. Only the qualitative analysis of  RQ$_2$, taking into account developers' discussion, can highlight the rationale for refactoring actions.}

\textbf{Conclusion validity.} In RQ$_1$ we performed a careful preprocessing of data \revised{and variable selection} to avoid multi-collinearity, and normalized metrics to allow properly interpreting the ORs.

\textbf{External validity.} Our analysis is limited to a sample of \systems Java open source projects hosted on GitHub, and the qualitative analysis to \manually PRs. We do not claim the generalizability of our findings to other programming languages or to industrial systems. For this reason, a further investigation on a more diverse set of projects, developed with different programming languages and belonging to both open and closed-source is highly desirable. \revised{Also, it is worth mentioning that in our manual analysis we only considered PRs for which \reftool identified at least one refactoring operation. This means that we did not consider PRs that, for example, targeted a complete remodularization of the system that involved refactoring operations not captured by \reftool.}

% !TEX root = main.tex

%%%%%%%%%%%%%%%%%%%%%%
%%%%%%%%%%%%%%%%%%%%%%
\section{Related Work} \label{sec:related}
%%%%%%%%%%%%%%%%%%%%%%
%%%%%%%%%%%%%%%%%%%%%%
%Several works in the refactoring field have proposed (semi-)automatic techniques to identify refactoring opportunities (see \eg \cite{Moha:tse2010,Palomba:tse2015}) and to recommend refactoring solutions (see \eg \cite{Tsantalis:tse2009,Bavota:tosem2014}). Given the goal of our paper, we remind the interested reader to surveys in the field for a more complete overview of these techniques \cite{Mens:tse2004,Bavota:refSurvey,Palomba:survey}. 

Different studies have empirically analyzed refactoring operations from different perspectives, including how developers perform refactoring \cite{Emerson:tse2011}; the relationship between refactoring and other software-related activities (\eg merge conflicts \cite{Mahmoudi:saner2019}); the impact of refactoring operations on the likelihood of introducing bugs \cite{Bavota:scam2012}; the impact of refactoring on specific quality indicators (\eg quality metrics) \cite{Stroggylos:2007,szoke2014bulk,Alshayeb20091319,Chavez:sbes2017}, or on developers' productivity \cite{Moser:2008}. While these studies mine and analyze refactoring for different purposes, they address different research questions as compared to the ones subject of our work. For this reason, we focus the discussion on studies analyzing why developers perform refactoring.

Wang \etal \cite{Wang:icsm2009} interviewed 10 developers from four software companies to reveal the major factors motivating refactoring operations.
% From the data collected during interviews, the authors build an empirical model featuring twelve \emph{intrinsic} and \emph{external motivators}. 
Results highlight external motivators \eg \emph{Recognitions from Others}, and  intrinsic motivators, \ie when refactoring is initiated without any obvious external reward (\eg \emph{Self Esteem}) behind refactorings.

Kim \etal \cite{Kim:fse2012} present a field study surveying and interviewing 328 Microsoft engineers to investigate when and how engineers refactor code. They identify the low readability of source code as the most important symptom that pushes developers to perform refactoring (mentioned by 21\% of developers). Our quantitative analysis confirms the central role of low code readability in triggering refactoring. Other frequently mentioned symptoms were code duplication (11\%), and fostering code reuse (10\%).

Silva \etal \cite{Silva:fse2016} observe that the previously discussed works \cite{Wang:icsm2009,Kim:fse2012} report findings from surveys asking developers about their general refactoring habits, without focusing on real refactorings they performed. Thus, they use \reftool \cite{Tsantalis:icse2018} to monitor refactorings performed in open source repositories and contacted the developers authoring the refactorings asking to motivate the performed changes. Then, they grouped the responses and defined a catalog of 44 motivations for 12 refactoring operations. For example, they find that three main motivations exist for \emph{Rename Package} refactoring: improve the package name, enforce naming consistency, and move package to appropriate container \cite{Silva:fse2016}. 

While we share with previous work \cite{Wang:icsm2009,Kim:fse2012,Silva:fse2016} the goal of identifying factors motivating developers to perform refactoring operations, we adopt a different and complementary experimental design. First, we quantitatively investigate process- and product-related factors that may correlate with refactoring actions (\eg does low code quality as assessed by quality metrics trigger refactoring operations?). Second, we qualitatively analyze refactorings performed by developers in the context of \manually merged PRs, to complement and validate the refactoring motivators already identified in the literature \cite{Wang:icsm2009,Kim:fse2012,Silva:fse2016}. Note that, as for the work by Silva \etal \cite{Silva:fse2016} and differently from the ones by Wang \etal \cite{Wang:icsm2009} and Kim \etal \cite{Kim:fse2012}, we look at motivations related to specific refactoring operations (\ie those performed in the analyzed PRs). Differently from Silva \etal \cite{Silva:fse2016}, we do not rely on answers collected through a survey, but we inspected the code changes performed in the commits related to the PRs, and read the reviewing process carried out before merging the PR and the related discussion. %This allowed us to get a complete picture of the context in which the refactoring has been performed and of the motivations behind it. 
As previously discussed, we validated and complemented the taxonomy of motivations they defined.

Peruma \etal \cite{Peruma:iwor2018} also look at the motivations pushing developers to refactor their code, but they focused only on rename refactoring operations, finding that, in most cases, renaming is applied to narrow the identifier meaning. Our qualitative study confirmed this finding.

Bavota \etal \cite{Bavota:jss2015} study the relationship between metrics, code smells and refactoring. Their findings indicate that there is no clear causation between code having a smell or exhibiting elevated complexity metric values and subsequent refactoring of this code. Our study, performed at the commit level, confirms their results but also points out that readability metrics and process-related factors play a significant role.
%Our quantitative analysis confirmed this finding. 

%As compared to our work, Bavota \etal \cite{Bavota:jss2015} detect refactorings at release level (\ie between two subsequent releases $r_i$, $r_{i+1}$). Then, they correlate these activities with the code quality proxies (\ie quality metrics and code smells) as measured/detected on $r_i$. This process is likely to introduce a substantial bias since, a file $F_k$ refactored in the time period between $r_i$ and $r_{i+1}$ might have been subject to refactoring activities close to the release of $r_{i+1}$ and may be substantially different as compared to its $r_i$'s version, on which the authors measured quality metrics and detected code smells. In our study, we adopt a much more precise approach, detecting refactorings in a specific commit $c_i$ and measuring quality metrics and code smells on the impacted files in their version right before the changes introduced by $c_i$.

Finally, very related work in this line of research is the one by Vassallo \etal \cite{Vassallo:scp2019}, in which the authors mine 200 systems to quantitatively investigate factors correlating with refactoring. They consider, in isolation, factors related to when, why, and by whom refactoring is performed.
%They consider: (i) time-related factors, \ie whether the refactoring is performed in the project startup phase or close to a release; (ii) the goal of the commit implementing the refactoring as identified by analyzing the commit message (\ie enhancement, new feature, bug fixing, or refactoring); and (iii) developer-related factors, such as the developers' workload, code ownership, and newcomer status. 
They found that (i) refactorings are mostly performed after one year from the project startup and rarely close to a new release; (ii) most of refactorings are performed while enhancing existing features; and (iii) developers that refactor code are often the owners of the impacted files. Our work, besides quantitatively investigating a more comprehensive set of process- and product-related factors (considered altogether in a mixed model), also complements this analysis with a qualitative investigation on the motivations behind refactorings.

\section{Conclusion} \label{sec:conclusion}
%%%%%%%%%%%%%%%%%%%%%%
%%%%%%%%%%%%%%%%%%%%%%
In this paper we quantitatively and qualitatively analyzed the reasons behind refactoring operations performed by developers. Our quantitative analysis highlighted that (i) code readability is the product-related factor mostly correlated with refactoring operations, and (ii) process-related factors such as source code change- and fault-proneness and, especially, the experience of developers changing a code component, play a significant role in triggering refactoring operations. Our qualitative analysis resulted in an extensive taxonomy of \taxonomycats motivations behind refactorings, relating to  quantitative results where possible. We have made the study material and data available in our replication package \cite{replication}.

The implications of our study trigger several directions for future work.

% \begin{itemize}

% \item 
\revised{\emph{Identifying ``when'' to trigger refactoring recommendations.} Our quantitative study (RQ$_1$) shed some light on the product and process-related factors that contribute to trigger refactoring operations. Such an empirical evidence represents the basis for future approaches able to predict  ``when'', during a project's evolution to trigger refactoring recommendations. Indeed, such an aspect is currently ignored in the refactoring recommenders literature, with researchers focusing their attention on the core problem of generating meaningful recommendations. However, recommending refactorings in a ``context'' in which developers do not feel in need to refactoring their code is unlikely to provide benefits. An interesting research direction in this field is to build on top of our RQ$_1$'s findings to devise models able to predict when refactoring recommendations would be welcome by software developers.}

% \item
\revised{\emph{On the relevance of ``semantic metrics'' for code smell detectors.} As shown in our RQ$_1$, the product metrics exhibiting the stronger relationship with refactoring operations are those exploiting textual information extracted from the code, usually referred to as semantic metrics. This was the case for the readability and for the C3 (Conceptual Cohesion of Classes) metrics. This result somehow supports the previous findings reported in the literature indicating a stronger alignment between the developers' perception of code quality and semantic metrics as compared to structural ones \cite{Bavota:icse2013}. These findings suggest the adoption of these metrics in code smell detectors that, at the end, are used to identify refactoring opportunities.}

% \item
\revised{\emph{On learning refactoring operations.} While discussing the results of our qualitative study (RQ$_2$), a bold message repeatedly emerged from the analyzed cases: the motivations behind a refactoring are variegated and heterogeneous and, probably, cannot be captured by any combination of metrics. This implies that refactoring recommenders using product- and/or process-metrics as fitness functions to recommend refactorings  will always miss many refactoring opportunities. Indeed, these techniques mostly target complex code components for refactoring (\eg God Classes in the case of extract class refactoring), ignoring all the other scenarios we documented (\eg applying extract class to remove a code clone). One possibility worth exploiting in the future is the application of deep learning techniques to refactoring recommenders. Indeed, recent work already shown the possibility to learn from code changes~\cite{Tufano:icse2019}. However, no previous work has attempted to design refactoring recommenders learning from developers' activities what a meaningful refactoring is in a given context.}

% \item
\revised{\emph{Automatic support for relevant refactorings is missing.} In RQ$_2$ we found many cases of manually performed refactoring operations that could benefit from the development of techniques to automate such code transformations. This includes the identification and automatic refactoring of redundant code and misuse of checked/unchecked exceptions in Java as well as the lack of support for preemptive refactoring (\ie refactoring operations aimed at accommodating future changes). We see these areas as possible directions for future work in software refactoring.}

% \item
\revised{\emph{Lack of production-ready tools.} Some of the cases discussed in RQ$_2$ highlighted that, while the research community has developed many good approaches for supporting specific refactorings, these approaches find little application in practice. One clear example of this are the rename refactoring operations suggested by reviewers to expand abbreviations used in the identifiers of the code contributed in PRs. These recommendations could be easily generated at commit time by one of the many approaches proposed in the literature for the automatic expansion of abbreviations (see \eg \cite{Lawrie:ICSM2011}). However, the lack of production-ready tools could be the reason for the lack of adoption of such techniques. This is an opportunity not only for researchers, but also for developers interested in building tools useful for the partial automation of code review activities.}

% \item 
\revised{\emph{On the need for pursuing ``trade-offs'' when refactoring.} Finally, our RQ$_2$ also highlighted the need for refactoring techniques able to consider the many contrasting objectives that a code transformation brings with it. Indeed, while most of the refactoring recommenders strive to maximize maintainability, our findings show that maintainability is only one of the many aspects considered by developers while refactoring, accompanied by performance, testability, \etc Future approaches to support refactoring should consider the pros and cons of the recommended solutions from different perspectives, suggesting operations that are sensible to different quality criteria.}

% \end{itemize}

Based on the above findings, our future research agenda will focus on the two points described above: (i) predicting \emph{when} to recommend refactorings, and (ii) developing refactoring tools sensible to different non-functional requirements.

\section*{Acknowledgment}
Pantiuchina thanks the Swiss National Science foundation for the financial support through SNF Project JITRA, No. 172479. This project has received funding from the European Research Council (ERC) under the European Union's Horizon 2020 research and innovation programme (grant agreement No. 851720).

\bibliographystyle{ACM-Reference-Format}
\bibliography{main}

\end{document}